\documentclass[twocolumn,floatfix,english]{revtex4}
\usepackage{pifont}
\usepackage{dcolumn}
\usepackage{amsmath}
\usepackage{amssymb}
\usepackage{graphicx}
\usepackage{bm}
\usepackage[utf8]{inputenc}
\usepackage[T1]{fontenc}
\usepackage{color}
\usepackage{url}
\usepackage[bf]{subfigure}
\usepackage{rotating}
\usepackage{mathrsfs}
 \usepackage{multirow}
 \usepackage{longtable}
 \usepackage{url}

\usepackage{lipsum, babel}

\begin{document}

\title{Scenarios for a post-COVID-19 world airline network}

\author{Jiachen Ye,$^{1,2}$ Peng Ji,$^{1,2*}$ Marc Barthelemy,$^{3,4}$}

\email{pengji@fudan.edu.cn, marc.barthelemy@ipht.fr}

\affiliation{
$^1$Institute of Science and Technology for Brain-Inspired Intelligence, Fudan University, Shanghai 200433, China\\
$^2$Potsdam Institute for Climate Impact Research (PIK), 14473 Potsdam, Germany\\
$^3$Institut de Physique Th\'eorique, Universit\'e Paris Saclay, CEA, CNRS, F-91191 Gif-sur-Yvette, France\\
$^4$Centre d'Analyse et de Math\'ematique Sociales, (CNRS/EHESS) 54, Boulevard Raspail, 75006 Paris, France
}

\date{\today}


\begin{abstract}

  The airline industry was severely hit by the COVID-19 crisis with an average demand decrease of about $64\%$ (IATA, April 2020) which triggered already several bankruptcies of airline companies all over the world. While the robustness of the world airline network (WAN) was mostly studied as an homogeneous network, we introduce a new tool for analyzing the impact of a company failure: the `airline company network' where two airlines are connected if they share at least one route segment. Using this tool, we observe that the failure of companies well connected with others has the largest impact on the connectivity of the WAN. We then explore how the global demand reduction affects airlines differently, and provide an analysis of different scenarios if its stays low and does not come back to its pre-crisis level. Using traffic data from the Official Aviation Guide (OAG) and simple assumptions about customer's airline choice strategies, we find that the local effective demand can be much lower than the average one, especially for companies that are not monopolistic and share their segments with larger companies. Even if the average demand comes back to $60\%$ of the total capacity, we find that between $46\%$ and $59\%$ of the companies could experience a reduction of more than $50\%$ of their traffic, depending on the type of competitive advantage that drives customer's airline choice. These results highlight how the complex competitive structure of the WAN weakens its robustness when facing such a large crisis.

\end{abstract}

\maketitle

\section{Introduction}

We see only the beginning of the socio-economic impact of the coronavirus pandemic (COVID-19) that spread over the whole world in the first semester of 2020. All sectors, agriculture, manufacturing industry, and of course the tertiary sector will be strongly affected by this crisis  \cite{nicola2020socio}, and our way of life could deeply change. In particular, the airline industry was severely hit with many governments that enforced both domestic and international travel restrictions at various degrees. Some countries restricted the flights from severely affected areas while others even cancelled almost all flights. These travel restrictions did delay or interrupt the further transmission of the COVID-19 \cite{chinazzi2020effect}, but also caused great damage to the world airline network (WAN) \cite{suzumura2020impact}, the most important travel network in today's world and one of the key infrastructures of today's global economy.

\begin{figure}[htbp]
  \centering
  \includegraphics[width=0.45\textwidth]{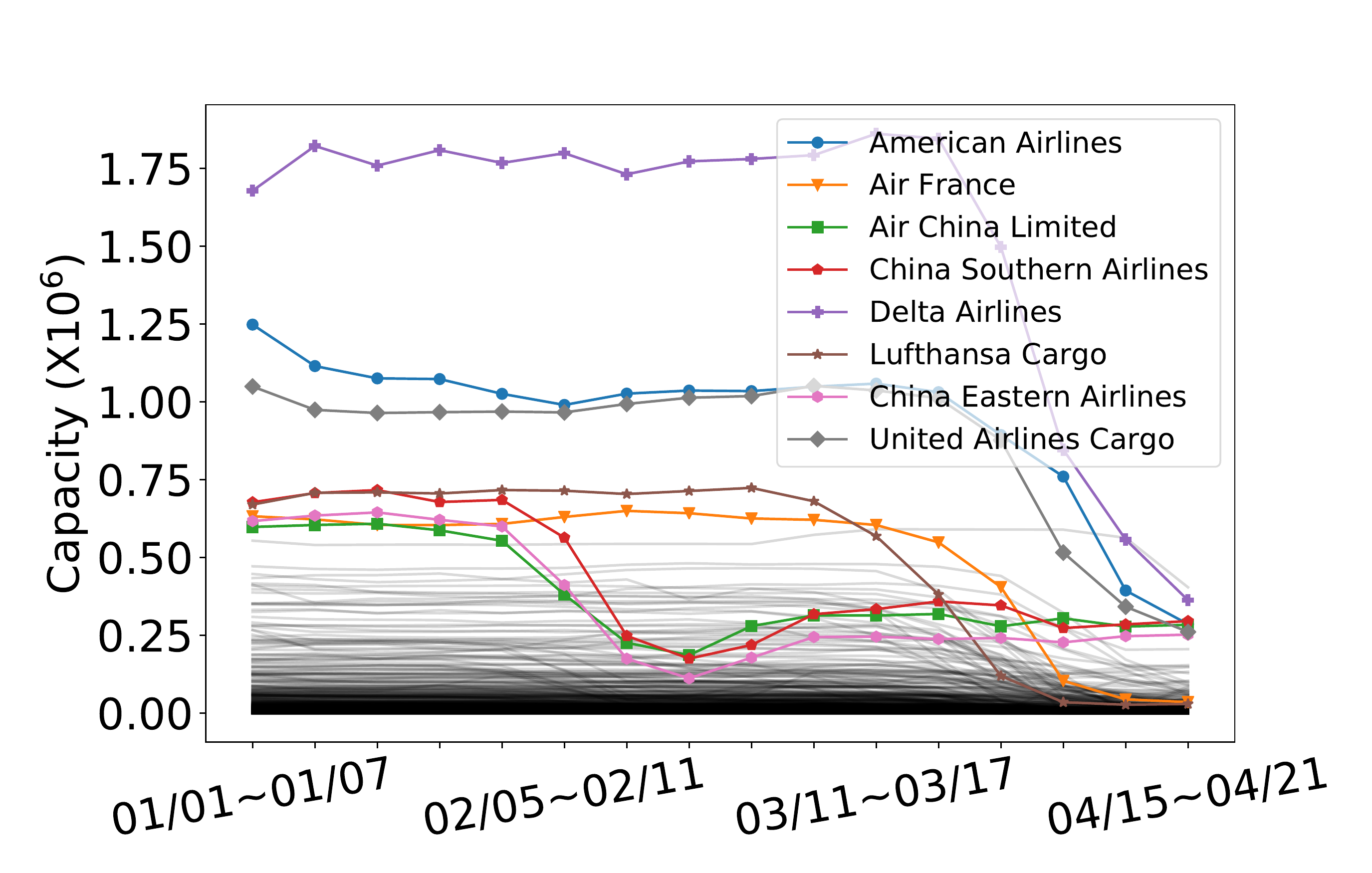}
  \caption{Evolution of the capacity of various airline companies early 2020. We highlight several large companies. The capacity of each company is averaged over every 7 days.}
  \label{ACN}
\end{figure}

More precisely, the COVID-19 outbreak caused a decline from $44,665$ segments in the period 1-7 January 2020 to $24,371$ in the period 19-25 April 2020 (representing a decrease of $45.4\%$), thus impacting a large number of companies. In order to visualize this variation, we average over a 7-day time window the capacity (given by the number of seats for all flights) for each company and plot the result on Fig.~\ref{ACN}. In particular, we show the capacity for the three biggest airline companies in China (Air China, China Southern Airlines and China Eastern Airlines) and how their capacity dropped around the beginning of February.
Many other airline companies also faced the decrease of capacity later with the epidemic spreading over the whole world, including Lufthansa Cargo at the end of February and Delta Airlines, American Airlines, United Airlines Cargo and Air France during the middle of March.

This is not the first time that the WAN suffered large-scale disruptions. For example the eruption of the volcano Eyjafjallajokull in 2010 not only threatened the safety of local residents, but also seriously affected the air traffic in Europe, by causing the cancellation of $108,000$ flights, disrupting the travel plans of $10.5$ million passengers and costing the airline industry in excess of $\$1.7$ billion in lost revenue \cite {budd2011fiasco, brooker2010fear}. There are, however, differences with these rare events, which brought massive losses for the air industry, and the current crisis. First, it remains unclear how long this pandemic will last. Since effect of high temperature and high humidity on COVID-19 seems to be limited \cite{wang2020high}, most experts do not expect that the epidemic will naturally come to an end in the summer 2020. As a consequence, governments may have to maintain the travel restrictions, which is definitely a strike to the whole air industry. Second, it remains unclear how the demand will resume back to normal, if it does so. An analysis of the Boston Consulting Group \cite{BCG} proposed various scenarios for the evolution of the air travel demand, the most probable one being a gradual recovery stretching into 2021 (`prolonged U-shape'). This is motivated by the fact that international travel is discouraged and that borders will slowly reopen leading to a slow return of the consumer confidence. The economic recession and the failure of travel distributors will enhance this drop of air travel demand. This situation could thus last a while, and has already provoked the bankruptcies of more than 20 airlines so far \cite{allplane,BI} such as Air Italy (Italy), Flybe (UK) which was the largest independent regional airline in Europe, Trans States Airlines (US), Compass Airlines (US), Virgin Australia (Australia), Avianca (Colombia), etc. Other major airlines called for multibillion bailout in order to survive \cite{Guardian}.

There are of course different reasons why an airline goes bankrupt \cite{SF}, but it ultimately boils down to the lack of cash to cover the airline's liabilities. An important ingredient is then the structure of the WAN and in particular how the different airlines, which compose it, are superimposed and interact with each other. Another crucial ingredient is how customers will choose an airline rather than another one, and it is in these extreme cases that a competitive advantage will prove to be fundamental. In this work, we will address two aspects. First, we will analyze the robustness of the WAN against the bankruptcy of different airlines, and which failures are the most dangerous for its connectivity. In this part we will also study how the traffic can reorganize itself. In a second part, we will consider how the global traffic demand reduction will affect different airlines, and for this we will test various strategies of how customers choose their airline company.
For these studies, we will use real traffic data from the Official Aviation Guide (OAG) for the WAN (see Material and Methods).

\section{The effect of bankruptcies on the world airline network}

\subsection{Robustness and the `airline company network'}

Usually, the WAN is defined as the network composed of nodes that are airports and links that represent direct flights between two airports \cite{barrat2004architecture,guimera2005worldwide,zanin2013modelling}. An obvious fact but often ignored in many studies about the WAN is that it results in fact from the superimposition of routes belonging to different airlines, leading to the well-known emergent hub-and-spoke structure \cite{bryan1999hub}. Each airline optimizes its own profit and the WAN can thus be seen as the emerging network resulting from the choices of all these operators. Despite this important aspect, the structure and the robustness of the WAN have mostly been studied from a topological point of view, with a focus on the effect on the giant connected component of node or link removal \cite{dall2006vulnerability,lordan2014robustness,verma2014revealing}. However, as noted above, describing the WAN as a simple topological network might be oversimplified, and in order to reach realistic conclusions, it seems necessary to take into account other aspects of this network (for a review and research agenda about the robustness of this network, see \cite{lordan2014study}). In particular, some works studied this important airline aspect, and considered the WAN as a multilayer network where each layer is a different airline company \cite{zanin2013modelling, boccaletti2014structure, kivela2014multilayer,cardillo2013modeling}. We note that in \cite{cardillo2013modeling} the authors considered an interesting study of re-scheduling of passengers after the failure of a given segment.

Here, we will analyze the robustness of the WAN when facing airline bankruptcies. This is different from the usual robustness study where nodes/links are removed. Instead, the removal of an airline implies the removal of possibly many segments, and/or the decrease of capacity if the segment is shared with another company, or the complete removal of the link if the airline was the only one operating on it. We first construct the WAN based on the data provided by the OAG for the period 1st January - 25th April 2019 (see Material and Methods for details). In order to analyze the robustness of this multilayer network from this point of view, we then construct what we call here the `airline company network' (ACN) where nodes represent airline companies (the number of nodes is here \(m=849\)), and two nodes in this network are connected if the corresponding companies share at least a segment. Obviously, two airline companies can share more than one segment and it is natural to define the weight of a link in the ACN as the segment overlap given by
\begin{equation}
O_{ij} = \left\{
\begin{aligned}
&\sum_{\ell \in L_i \cap L_j} \frac{\min \{N_i(\ell),N_j(\ell)\}}{\min \{N_i,N_j\}}~~~&i \neq j\\
&~~~0 &i=j
\end{aligned}
\right.
\end{equation}
where \(L_i\) is the set of segments operated by company \(i\), \(N_i(\ell)\) is the capacity of company \(i\) on segment \(\ell\) (averaged over \(T=116\) days) and \(N_i=\sum_{\ell \in L_i}N_i(\ell)\) is the total capacity of company \(i\). This segment overlap captures the similarity between two airline companies and its distribution is shown in Fig. S1 in the appendix. We observe a wide range of this overlap and while most of the companies do not share segments with each other, nearly 100 pairs of companies display an overlap equal to 1, indicating that either one is included in the other, or that they are  proposing exactly the same set of segments and are thus in direct competition.

The study of this weighted ACN allows us to have a better view of how the different airlines interact with each other. In particular, it will help us to understand how robust the WAN is when airlines are failing. In general, when studying the robustness of the WAN, different link or node removal strategies are adopted \cite{dall2006vulnerability,lordan2014robustness}. Removing a link implies that all companies operating on this segment stopped simultaneously, which is unlikely to occur in general, except in cases of airport closures which happened for example during the eruption of the volcano Eyjafjallajokull. Here, we will consider the specific situation where many airlines can fail. The bankruptcy of an airline does not imply, however, that one or more segments in the WAN will disappear, but just that its capacity will be reduced.

Based on the segment overlap, it is natural to define the average segment overlap of a company with its `neighbours' as
\begin{equation}
O_i = \frac{1}{|C_i|}\sum_{j \in C_i} O_{ij},
\end{equation}
where \(C_i=\{j~|~O_{ij}>0\}\) is the set of companies sharing at least one segment with company \(i\) and $|C_i|$ represents the number of corresponding companies. A large average segment overlap indicates that the corresponding airline company shares many segments with others and is thus in direct competition with them. In contrast, a small overlap indicates a company that operates alone on some segments and does not suffer any competition there.  It is then tempting to think that the average segment overlap can serve as an indicator for identifying `critical' companies, whose bankruptcy will induce the isolation of certain airports and cities and cause a great damage to the connectivity of the WAN.

Classically, when ignoring the structure in terms of airline companies, the most effective strategy to destroy the connectivity of the WAN is to remove the most central nodes (i.e. nodes with highest betweenness centrality, see \cite{dall2006vulnerability,lordan2014robustness}) or the links with the lowest traffic \cite{verma2014revealing}. As already mentioned above, these results do not provide much information about the impact of airline bankruptcies and the partial reduction in traffic capacity for certain segments. In order to test this influence, we first consider the robustness of the WAN for four different removal strategies: (i) random removal of airline companies; (ii) removal of airline companies with the largest passenger capacity \(N_i\); (iii) removal of airline companies with the largest betweenness centrality computed in the weighted ACN; (iv) removal of airline companies with the largest average segment overlap \(O_i\). We note that for strategy (iii), the calculation of the betweenness centrality in the weighted ACN relies on the shortest path between two nodes such that the sum of weights (here the segment overlap $O_{ij}$) is maximal among all connective paths. As most robustness analysis do, we choose the size of the giant component (i.e. the connected subgraph with the largest number of nodes), as the measure of the connectivity of the WAN.

\begin{figure}
  \centering
  \includegraphics[width=0.42\textwidth]{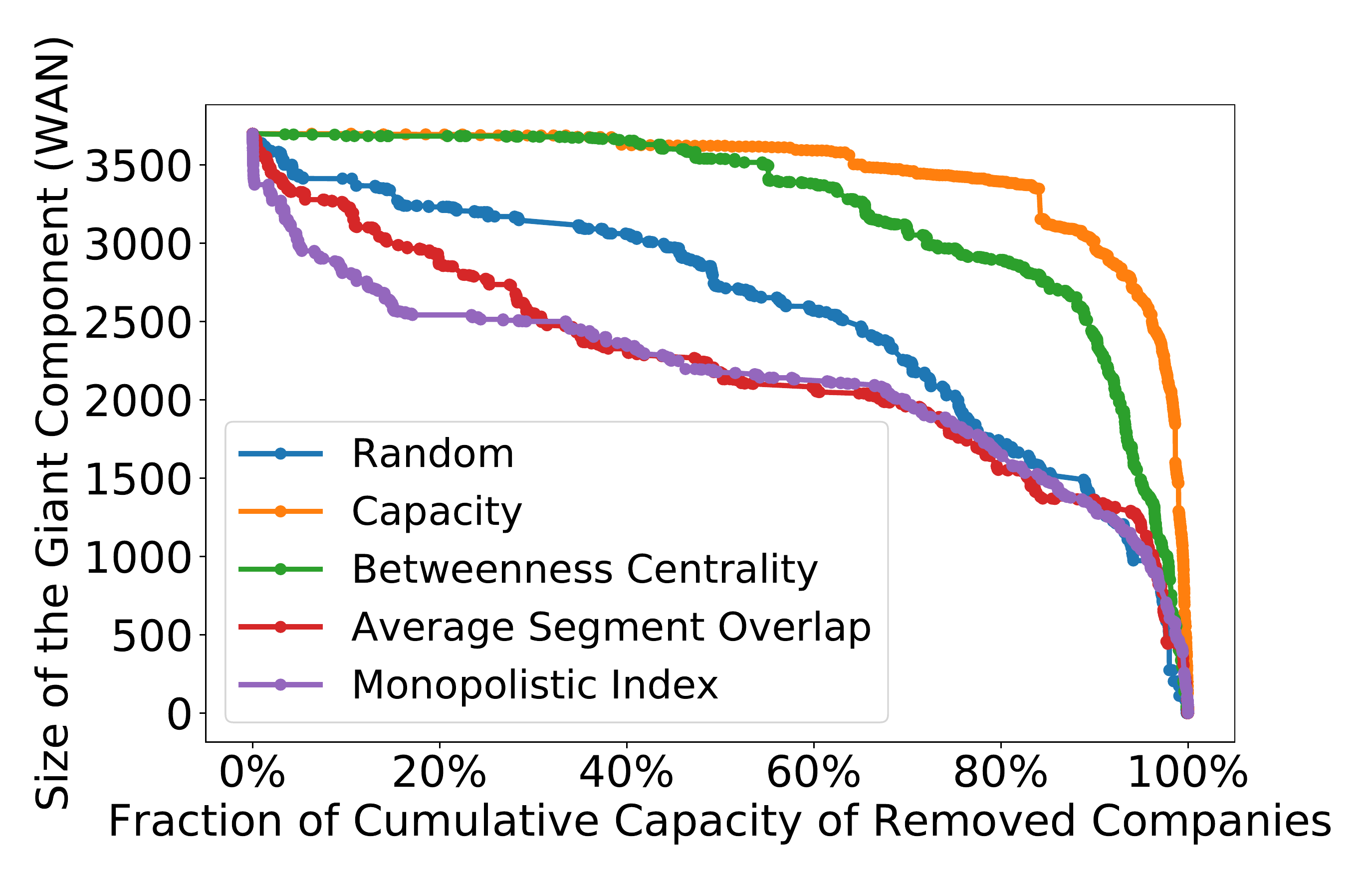}
  \caption{Size of the giant component of WAN as the function of the fraction of the cumulative passenger capacity of removed companies. Five different removal strategies of airline companies are compared.}
  \label{Robustness}
\end{figure}

We show in Fig.~\ref{Robustness} the size of the giant component of the WAN as a function of the fraction of the cumulative passenger capacity of removed companies (which implies that the intervals along the x-axis between dots are not uniform due to the diversity of the companies' capacities).
For the strategy (ii), where we remove companies according to their passenger capacity (from the largest to the smallest), the variation over the x-axis is large but the impact on the WAN is small. Sorting and removing companies according to their betweenness centrality in the ACN is also not very effective. Compared to these cases, removing airline companies in a random order has actually a larger impact on the WAN structure. However, we observe that the last strategy (iv) is very effective: removing companies according to their integration with other companies seems to have the largest impact on the WAN's connectivity, suggesting that crucial paths in the WAN are shared by the same set of companies. More generally, this is a sign that in order to understand the robustness of the WAN and the impact of the failure of an airline company, we need to take into account its interaction with the other companies.

\begin{figure}
  \centering
  \includegraphics[width=0.42\textwidth]{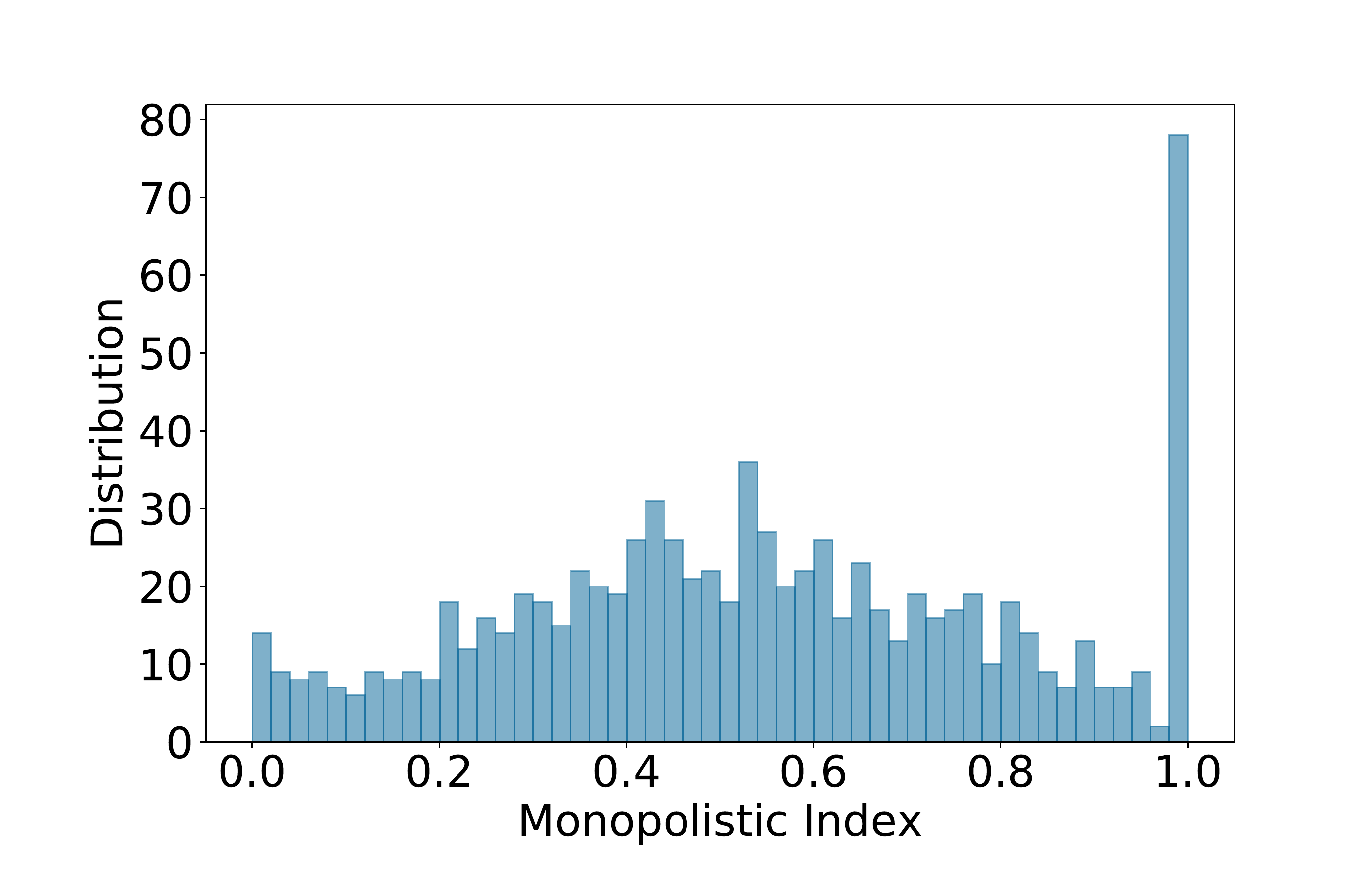}
  \caption{Distribution of monopolistic index.}
  \label{Monopoly}
\end{figure}

This discussion shows that not all airlines are equivalent and that their bankruptcy can display a large variety of consequences. In particular, the location of the airline in the ACN, or in other words, how it is coupled to other companies, seems to be a crucial ingredient for understanding the impact of its bankruptcy. This leads us to quantify how much a company has the monopoly over its different segments and we define the monopolistic index as
\begin{equation}
M_i = \frac{1}{\vert L_i \vert}\sum_{\ell \in L_i} \frac{N_i(\ell)}{\sum_{j=1}^m N_j(\ell)},
\end{equation}
where $|L_i|$ is the total number of segments serviced by company  $i$. This index is equal or close to 1 if most the segments of company \(i\) are dominated by \(i\). The distribution of $M_i$ shown in Fig.~\ref{Monopoly} indicates that there is a large variety of companies with various degree of monopoly, demonstrating the large diversity of the airlines companies. The peak at $M=1$ indicates that there are about 80 companies which are totally monopolistic: all their segments are not shared with another company.

The importance of this feature can be checked with our robustness test: we remove companies according to their monopolistic index and obtain the result shown in Fig.~\ref{Robustness}. The rapid decrease of the WAN connectivity with this strategy confirms our conclusion about the importance of airlines interactions for understanding the structure of the WAN and its robustness.

\subsection{Rerouting after bankruptcy}

An important process not considered in the robustness analysis above is the direct consequence on passengers of an airline company bankruptcy. These passengers have to be rerouted and need to choose flights of other companies (when it is possible), putting extra stress on some flights. In some cases even, the bankruptcy of companies can isolate some cities leaving no other choices for the passengers to find another transportation mode.  In this section, we will simulate this process for various level of demand decrease, starting from the real data, in order to understand the possible post-COVID-19 scenarios for the WAN.

We denote by \(\alpha \in [0,1]\) the average demand/load ratio for all flights. We consider here the uniform case (see the next section for other assumptions), where the actual number of passengers for airline company \(i\) on segment \(\ell\) is then \(\alpha N_i(\ell)\), where the capacity is $N_i(\ell)$. If the company $i$ fails,  its $\alpha N_i(\ell)$ passengers have to be rerouted for each of its segment $\ell$. We assume here that passengers always choose the shortest path with less transfers as possible. If for the segment $\ell$, there are other direct flights provided by other companies, some lucky passengers can take them (which will be possible with $\alpha<1$). Once the flights of current shortest path are fully occupied, the rest of the passengers will be rerouted over the second shortest path, and so on.  If all the rerouted passengers can reach their final destinations, we compute the average number of additional transfers and the average additional distance as the `cost' of the airline's bankruptcy.
For example, if a passenger wanted to fly from city \(a\) to \(b\) at first, now he/she has to transfer in city \(k\). In this case, his/her number of additional transfers is 1 and the additional distance is \((d_{ak}+d_{kb})-d_{ab}\), where \(d_{ab}\) is the distance between city \(a\) and \(b\). In the less favorable case where some passengers cannot reach their destination which can happen if all flights are fully booked, the WAN fails and we compute the number of these stranded passengers.

We test the bankruptcy of each airline company in the database, and use the rerouting process as described above. We show in Fig.~\ref{Rerouting} the weighted average cost of additional transfers with weight equal to the capacity of companies not inducing stranded passengers (see Fig.~S2(a) for the additional distance). For a low global demand, passengers can always be rerouted without too many problems, but with increasing $\alpha$, passengers need to make always more transfers and at the cost of a larger distance.
\begin{figure}
  \centering
  \includegraphics[width=0.40\textwidth]{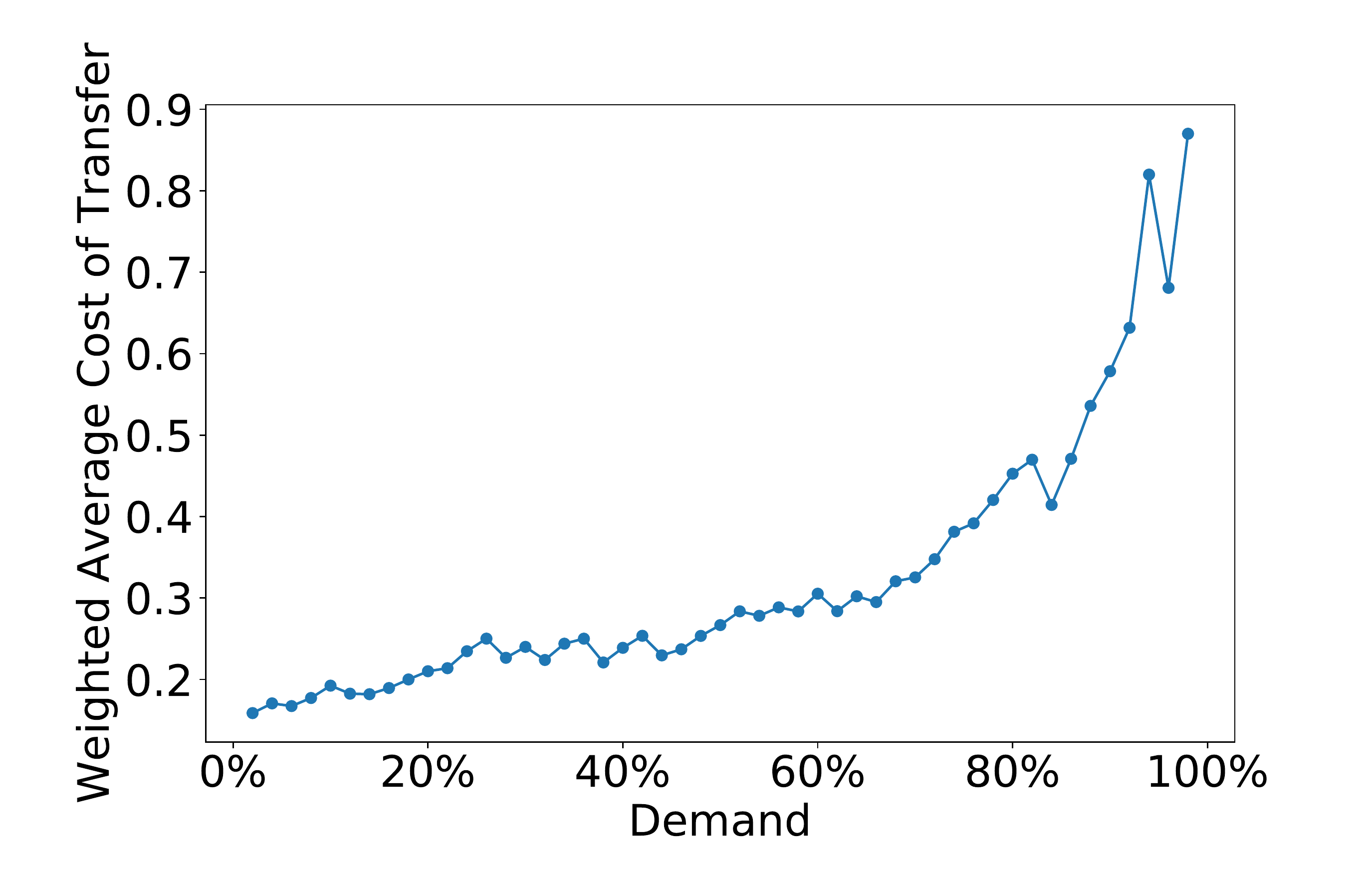}
  \caption{Weighted average cost of transfers number due to rerouting (with weight given by the capacity of companies).}
  \label{Rerouting}
\end{figure}
We also consider the total number of passengers stranded due to the bankruptcies of airlines if they cannot be successfully rerouted (Fig.~S2(b)). As an indication, we fit the result with a power law function and obtain the large exponent value of $3.75$, indicating a rapid increase with $\alpha$ of the number of stranded passengers: when \(\alpha\) is of the order of $0.5$, there is a significative number of stranded passengers of about 2.5 millions.

The existence of stranded passengers depends on a critical point of demand for each airline company (denoted by $\alpha_i^c$ for company $i$) above which its passengers cannot transfer successfully after its bankruptcy, or in other words: bankruptcy of company \(i\) leads to stranded passengers only for $\alpha>\alpha_i^c$. The distribution of this demand threshold is shown in Fig.~S2(c). We observe a roughly uniform distribution for all values of $\alpha^c_i$ and a peak at $\alpha^c_i=0$ which corresponds to the case of cities served by a single company. When these companies go bankrupt, passengers have no other means to reach their destination (this is the case for approximately 300 companies). We also observe a peak around $\alpha^c_i\approx 1$ signaling the existence of very robust situation for which their bankruptcy almost never lead to stranded passengers.

\section{The effect of a lower demand and customer's airline choice strategies}

When the global demand decreases, we have assumed so far that all segments and all companies are affected in the same, uniform manner.
This is, however, not true in general, and we have to test various customer's strategies for choosing a company. A reduction in the global demand can thus have different local consequences and lead to very low passenger traffic demand for some companies. These are obviously the companies that will be the most at risk and it is the focus of this part to identify them.

We first focus on one segment \(\ell\) shared by multiple companies. Before the crisis, each company has its capacity \(N_i(\ell)\) on this segment and the total number of seats is \(N(\ell)=\sum_i N_i(\ell)\). We assume that the global demand is still described by $\alpha\in [0,1]$ for all segments (for further studies, we could assume some geographical dependence), the actual number of passengers on this segment is \(\alpha N(\ell)\). The question is then how we remove the $(1-\alpha)N(\ell)$ passengers on this segment $\ell$, or equivalently, how the remaining $\alpha N(\ell)$ passengers choose their airlines.
There are many factors influencing the decision for choosing a specific airlines, including the frequency, reliability, price, etc. (see for example the study \cite{dolnicar2011key}). We decided here to focus on two main aspects: the capacity of an airline and its rank. A large capacity usually goes together with a large frequency and reliability, and for the rank we use the rank list of world's top 100 Airlines 2019 provided by Skytrax \cite{skytrax}.

We thus consider the following 6 strategies so that all \(\alpha N(\ell)\) passengers find their seats on the flights among these companies: (i) The first `Uniform' Strategy (UN) used as a reference: Each company receives a number of passengers proportional to its capacity. This random case ignores all competitive avantage but provides a benchmark for assessing the effect of other strategies. (ii) The `Biggest First' Strategy (BF): Passengers are assumed to always give preference to the company with the largest capacity (and then to the second largest one, etc.); (iii) `One-Uniform' Strategy (1UN): Passengers are assumed to choose first the company with the highest rank (provided by \cite{skytrax}). If no companies on the rank list operate on this segment, a random company will be chosen first. After flights of the first-choice company are fully occupied, the other companies receive a number of passengers proportional to their capacities; (iv) `One-Biggest First' (1BF): Passengers are assumed to choose the company with the highest rank first. If no companies on the rank list are available for this segment, a random company will be chosen. After flights of the first-choice company are full, passengers give then their preference to the company with the largest capacity; (v) `Rank-Uniform' (RUN): Passengers are assumed to choose companies according to the rank list. If flights of companies on the list are occupied, the other companies receive a number of passengers proportional to their capacities; (vi) `Rank-Biggest First' (RBF): Passengers are assumed to choose companies according to the rank list. If flights of companies on the list are full, passengers give then their preference to the company with the largest capacity.

\begin{figure}
  \centering
  \includegraphics[width=0.45\textwidth]{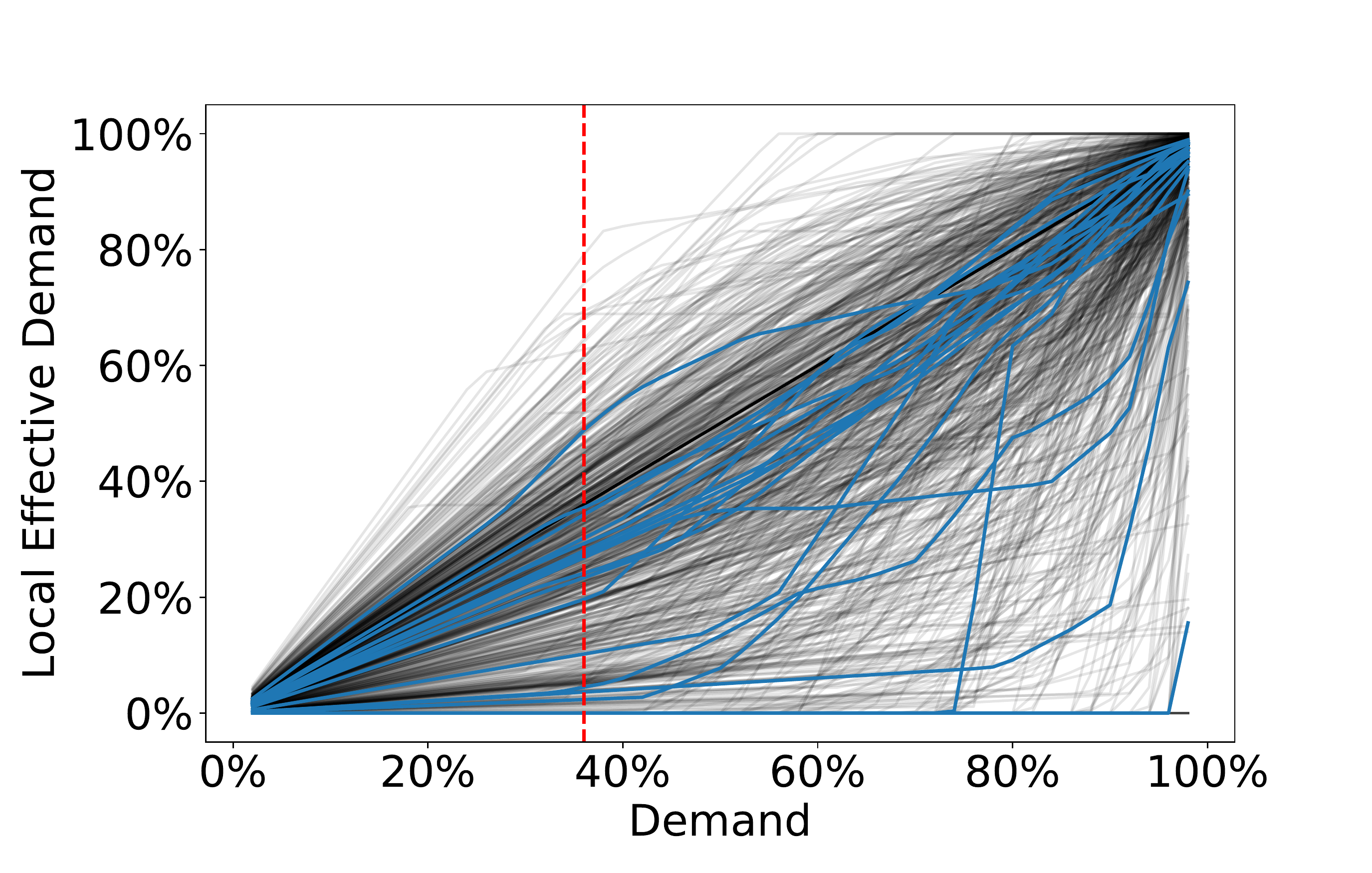}\\
    \includegraphics[width=0.45\textwidth]{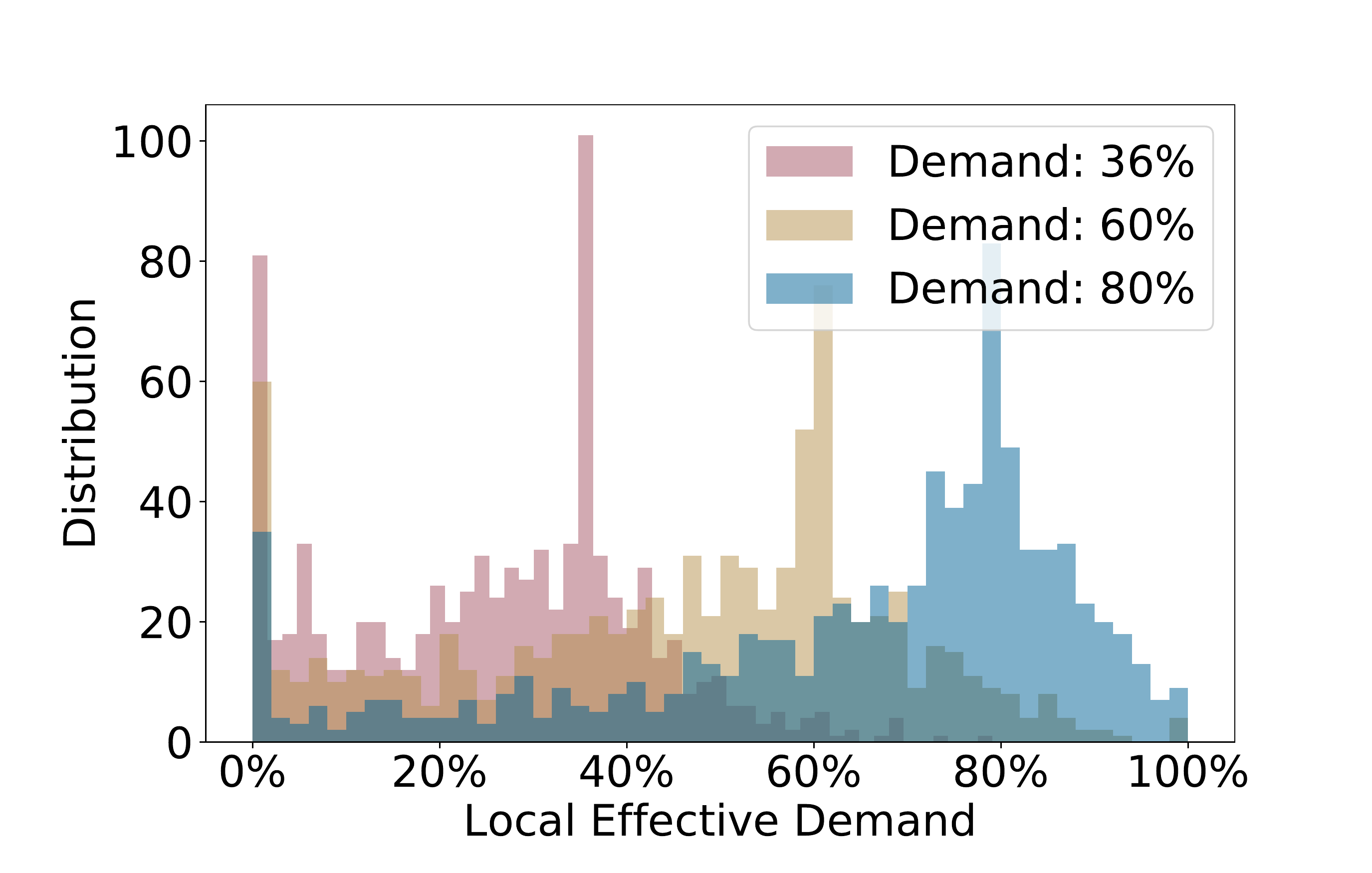}
  \caption{Local effective demand for the `Biggest First' strategy. (a) Change of local effective demand for all companies (we highlight the curves for the 23 companies that have gone broke). The vertical red dotted line represents the estimated value of global demand \(\alpha=36\)\% in April 2020. (b) Distribution of local effective demand for $\alpha=36\%$, $60\%$ and $80\%$.}
  \label{EffectiveDemand_BF}
\end{figure}

Considering all the segments for each company $i$, for a given strategy $s$ and for a given value of $\alpha$, we denote the actual number of passengers for company $i$ by $N^s_i(\alpha)$. The local effective demand for company $i$ (and for strategy $s$) is then given by
\begin{equation}
E_i^s(\alpha) = \frac{N^s_i(\alpha)}{N_i}.
\end{equation}
Companies with a large value of the local effective demand will have enough passengers to function normally, while those with a low local effective demand can be considered at risk of bankruptcy. For the strategy of `Biggest First', we show the change of local effective demand for all companies (Fig.~\ref{EffectiveDemand_BF}(a)), and its distribution when \(\alpha\) is equal to $36\%$ (the estimated value of demand in April 2020 according to IATA \cite{IATA}), $60\%$ and $80\%$ in Fig.~\ref{EffectiveDemand_BF}(b). Even if most companies have a local effective demand around $36\%$, there are about $80$ companies without any passengers. For  the 23 companies that went bankrupt so far, their local effective demand was in the range $[0\%,49\%]$, consistently with the idea that a company with demand less than $50\%$ is in financial danger.
\begin{figure}
  \centering
  \includegraphics[width=0.45\textwidth]{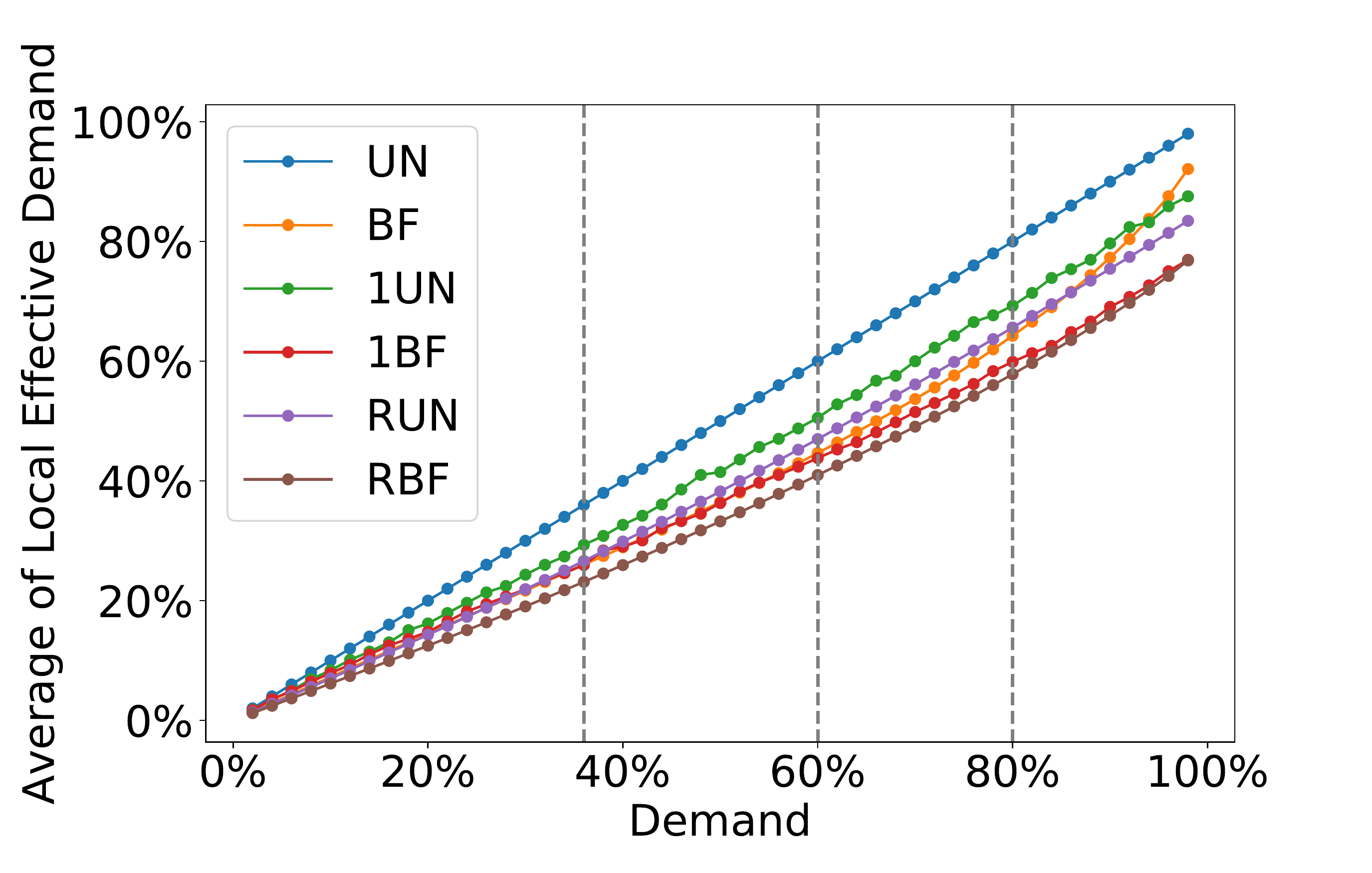}
  \includegraphics[width=0.45\textwidth]{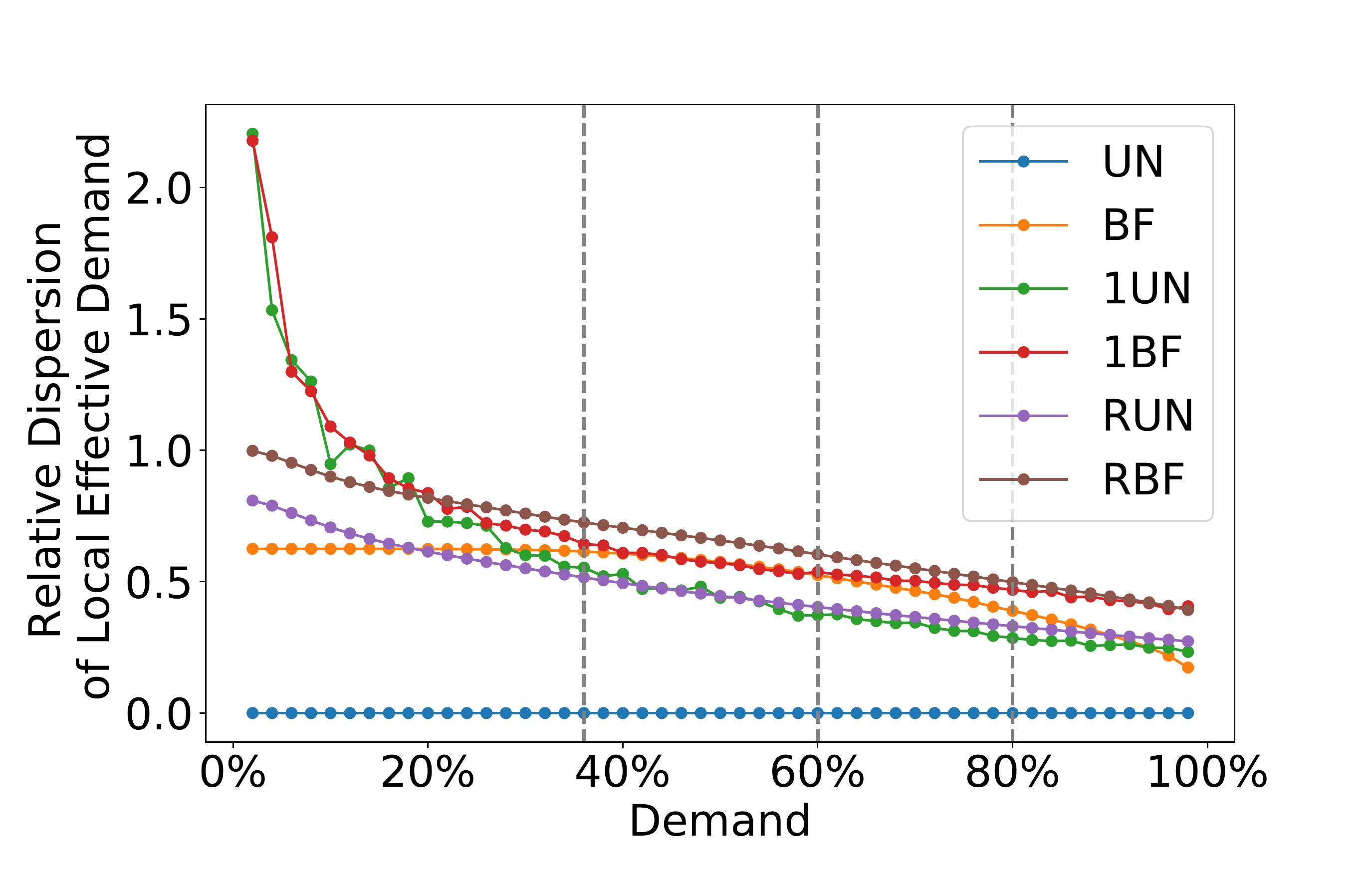}
  \caption{Comparison of local effective demand between 6 different strategies. (a) Average of local effective demand. (b)Relative Dispersion of local effective demand. The vertical gray dotted lines highlight $\alpha=36\%$, $60\%$ and $80\%$.}
  \label{EffectiveDemand_Comparison}
\end{figure}

\begin{figure}
  \centering
    \includegraphics[width=0.45\textwidth]{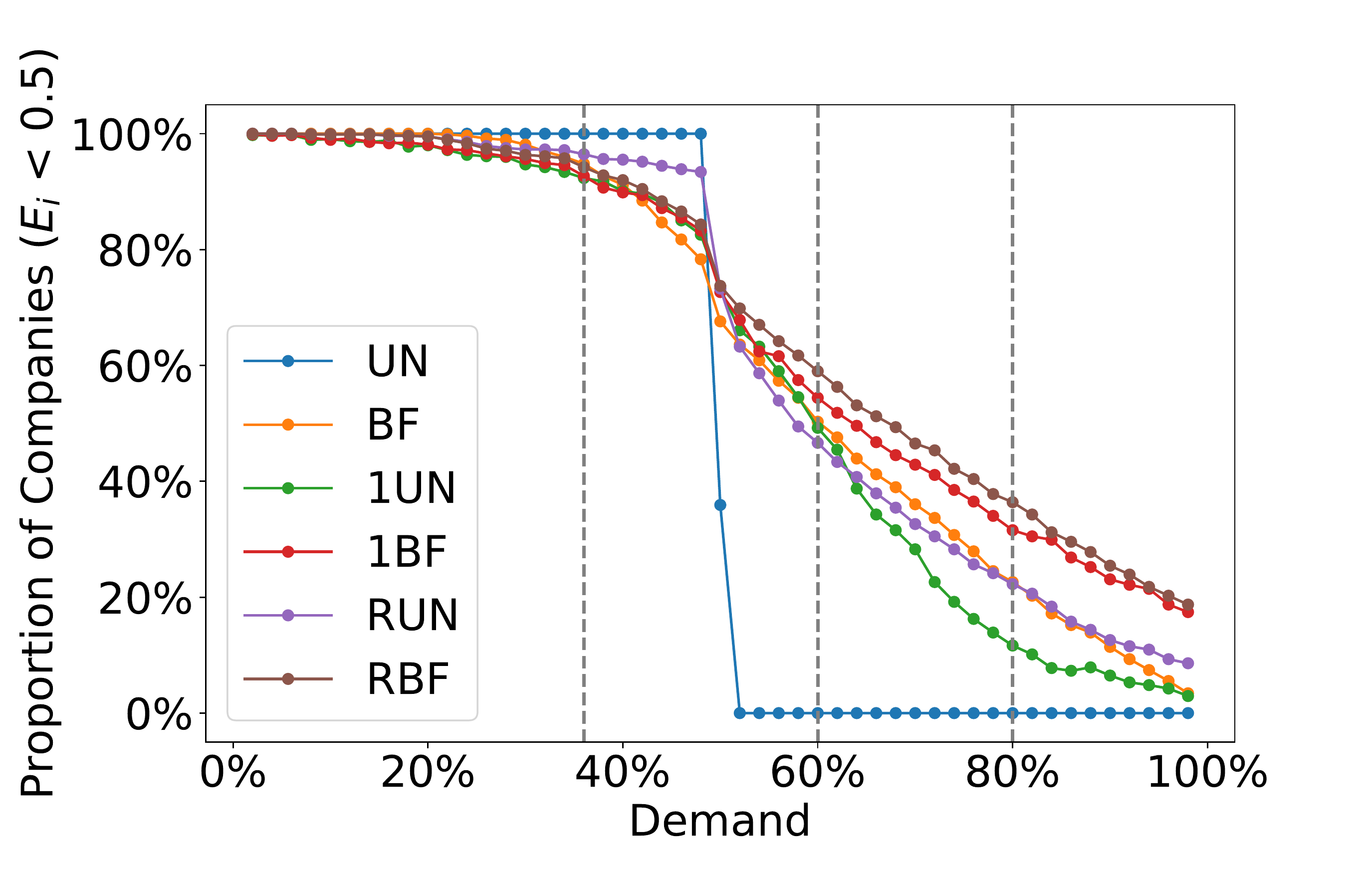}
  \caption{Proportion of companies with local effective demand less than $50\%$. The vertical gray dotted lines highlight $\alpha=36\%$, $60\%$ and $80\%$.}
  \label{Riskfraction}
\end{figure}

In order to compare quantitatively these different strategies, we show the local effective demand averaged over all companies and its relative dispersion (equal to the ratio of dispersion to average) in Fig.~\ref{EffectiveDemand_Comparison}(a,b). For the uniform case, where airlines are chosen completely at random, the local effective demand of each company is, as expected, identically equal to $\alpha$ ($E_i^{\text{UN}}(\alpha)\equiv \alpha$). All other strategies produce important variations: from Fig.~\ref{EffectiveDemand_Comparison}(a) we can roughly estimate that for a demand of $36\%$, certain companies can experience a local reduction of demand as low as about $30\%$ or even less if we take into account the dispersion which cannot be neglected (the relative dispersion is always of order 1 - see Fig.~\ref{EffectiveDemand_Comparison}(b)). Even for a global demand of $60\%$, all these strategies (except for the purely random case) predict that most companies will experience a decrease of $50\%$ or more of their traffic. In order to highlight this local heterogeneity effect, we plot in Fig.~\ref{Riskfraction} the proportion of companies with local effective demand less than $50\%$ (We choose here $50\%$ as this seems an already drastic reduction of the demand that can be barely sustainable for most companies. Results for other values of the local reduction give similar results, see Fig. S4). Except for the uniform strategy, Fig.~\ref{Riskfraction} shows that for the current state of air travel demand $\alpha=36\%$, more than $90\%$ of companies will see an actual number of passengers less than half of their capacity in the pre-crisis state. Even if the global demand is back to $60\%$, then between $46\%$ and $59\%$ of the companies will have an actual number of passengers less than half of their capacity in the pre-crisis state. In the unlikely event that it will go back in a near future to $80\%$, we still see from $11\%$ to $36\%$ of companies with half of their regular traffic.

\section{Discussion}

The results in this paper are twofold. First, the impact on the WAN connectivity of an airline company failure depends on its coupling with other companies. If the company shares many segments with others, its failure will more strongly affect the WAN structure. This shows that the impact of a bankrupted airline cannot be discussed independently from the other companies, and that it is necessary to take into account the whole structure of the coupling between companies - which is encoded in the ACN introduced here. Next, we showed that the global demand reduction can hide large heterogeneities in its impact at the airline level. More precisely, we have found that, depending on the customer choice strategy, the local reduction of demand can be extremely different from its average value, in particular in the presence of a competitive advantage. This shows that more companies than we can expect from a simple demand reduction argument could be strongly affected by this crisis. In the reasonable scenario where the global demand is back to $60\%$ of its total capacity, our analysis suggests that roughly half of the companies will see their traffic divided by two. These results need naturally to be enriched by many other details about companies, but we believe that they pave the way for constructing a robust tool allowing us to understand and predict the future of the WAN.

\section{Material and methods}

\subsection{Data availability}

The data that support the findings of this study were purchased at OAG
\url{https://www.oag.com/}.

\subsection{Data}

The Official Aviation Guide (OAG), one of the largest global travel data providers, provides the detailed information of every scheduled flight including the operating airline company, origin/destination airport, aircraft type, number of seats, departure date, etc.

We focus on the nonstop flights during the period from 1st January to 25th April 2020 for understanding the variation of global air traffic caused by COVID-19 and the corresponding period 1st January -- 26th April 2019 for constructing the world airline network (WAN) in `normal state'.  In the WAN, nodes represent cities and links denote existence of direct flights. Considering all the flights during 1st January to 26th April 2019, there are \(n=3,713\) cities serving as origins/destinations and \(n_\ell = 51,306\) different segments connecting these cities. As the comparison, there are $3,752$ cities and $49,215$ segments during the period 1st January to 25th April 2020. We define the weight of link as the everyday capacity (number of seats) of flights between cities, i.e.,
\begin{equation}
w_{ij} = \frac{1}{T}\sum_{t=1}^T \frac{N_{ij}(t) + N_{ji}(t)}{2},
\end{equation}
where \(N_{ij}(t)\) is the total capacity of flights from city \(i\) to city \(j\) on day \(t\), and \(T=116\) is the length of time range.


\bibliography{paper}
 \bibliographystyle{unsrt}


\onecolumngrid
\appendix

\setcounter{equation}{0}
\setcounter{figure}{0}
\renewcommand{\thefigure}{S\arabic{figure}}
\renewcommand{\theequation}{S\arabic{equation}}


\section{Segment overlap distribution}

For the airline company network (ACN), we define the segment overlap between companies as
\begin{equation}
O_{ij} = \left\{
\begin{aligned}
&\sum_{\ell \in L_i \cap L_j} \frac{\min \{N_i(\ell),N_j(\ell)\}}{\min \{N_i,N_j\}}~~~&i \neq j\\
&~~~0 &i=j
\end{aligned}
\right.
\end{equation}
where \(L_i\) is the set of segments operated by company \(i\), \(N_i(\ell)\) is the capacity of company \(i\) on segment \(\ell\) (averaged over \(T=116\) days) and \(N_i=\sum_{\ell \in L_i}N_i(\ell)\) is the total capacity of company \(i\). This quantity captures the similarity of segments operated by two different companies. The segment overlap $O_{ij}$ equal to 1 implies that either the set of segments of one company is included in the set of the other one, or that they operate on exactly the same segments. A value equal to 0 means that the corresponding companies operate over completely different sets of segments.

\begin{figure}[h]
  \centering
  \includegraphics[width=0.5\textwidth]{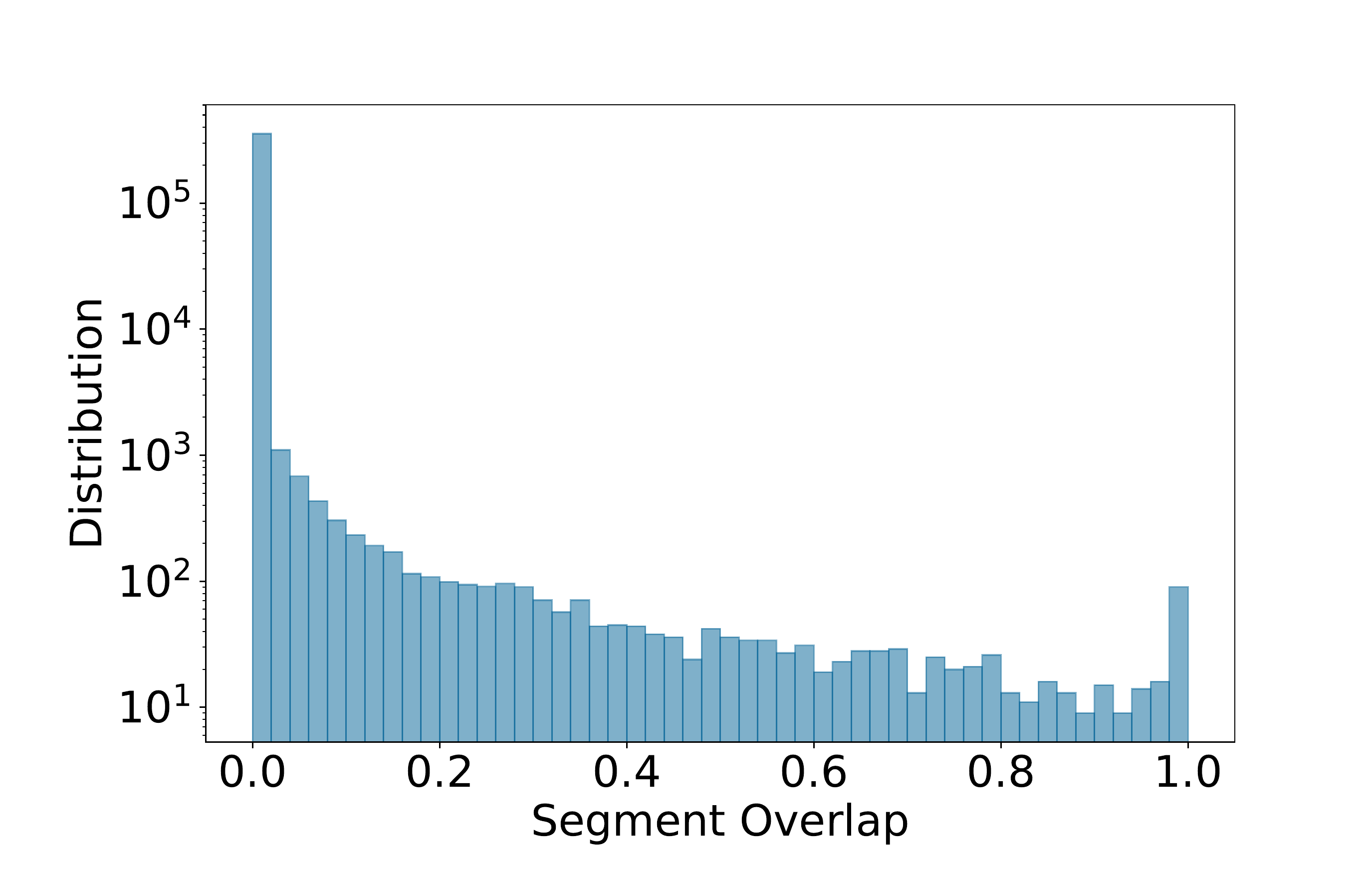}
  \caption{Segment overlap distribution for all companies (shown in linlog).}
  \label{fig:ACN}
\end{figure}

The distribution of segment overlap is shown in Fig.~\ref{fig:ACN}, and displays a wide range of values distributed in $[0,1]$. We also observe that while the majority of the companies do not share any segments with each other (the highest peak with segment overlap equal to 0), nearly 100 pairs of companies have an overlap equal to 1, meaning that they operate over the same set of segments (or the set of segments of a company is included in the set of the other one), indicating that these companies are in direct competition.

\section{Rerouting after bankruptcy}

For a given value of the demand \(\alpha \in [0,1]\), we compute the effect of rerouting the passengers. In particular, we estimate the additional distance experienced by all the rerouted passengers (the difference of distance between new and original routes), and count the number of stranded passengers.

For companies whose bankruptcy does not induce stranded passengers, we observe that passengers have to take always more transfers and have to travel over longer distance with the increase of \(\alpha\). Fig.~\ref{Rerouting}(a) shows the additional distance travelers take in order to reach their destination. When \(\alpha>70\%\), passengers have to travel more than a total of 1000 kilometers when transferring (results averaged over companies not inducing stranded passengers with weight given by the capacity).

For companies whose bankruptcy leads to stranded passengers, the total number of stranded passengers (denoted by \(N_s\)) is shown in Fig.~\ref{Rerouting}(b). When \(\alpha\) is of order of $0.5$, there is a significative number of passengers stranded of about 2.5 million. The exponent $\tau=3.75$ of the power law function fitting ($N_s\propto\alpha^\tau$) indicates the rapid growth of the number of stranded passengers with $\alpha$.

In Fig.~\ref{Rerouting}(c), we show the distribution of the critical demand threshold (for each company) above which there is a non zero number of stranded passengers. The two peaks of the demand threshold in Fig.~\ref{Rerouting}(c) correspond to two extremal types of airline companies: Companies with $\alpha^c_i\approx 0$ dominate the segments for at least one city and its bankruptcy will cause the isolation of the corresponding cities, while companies with $\alpha^c_i\approx 1$ have only a few passengers and their passengers will never experience difficulties to be rerouted after their bankruptcy.

\begin{figure}[h]
  \centering
  \includegraphics[width=0.45\textwidth]{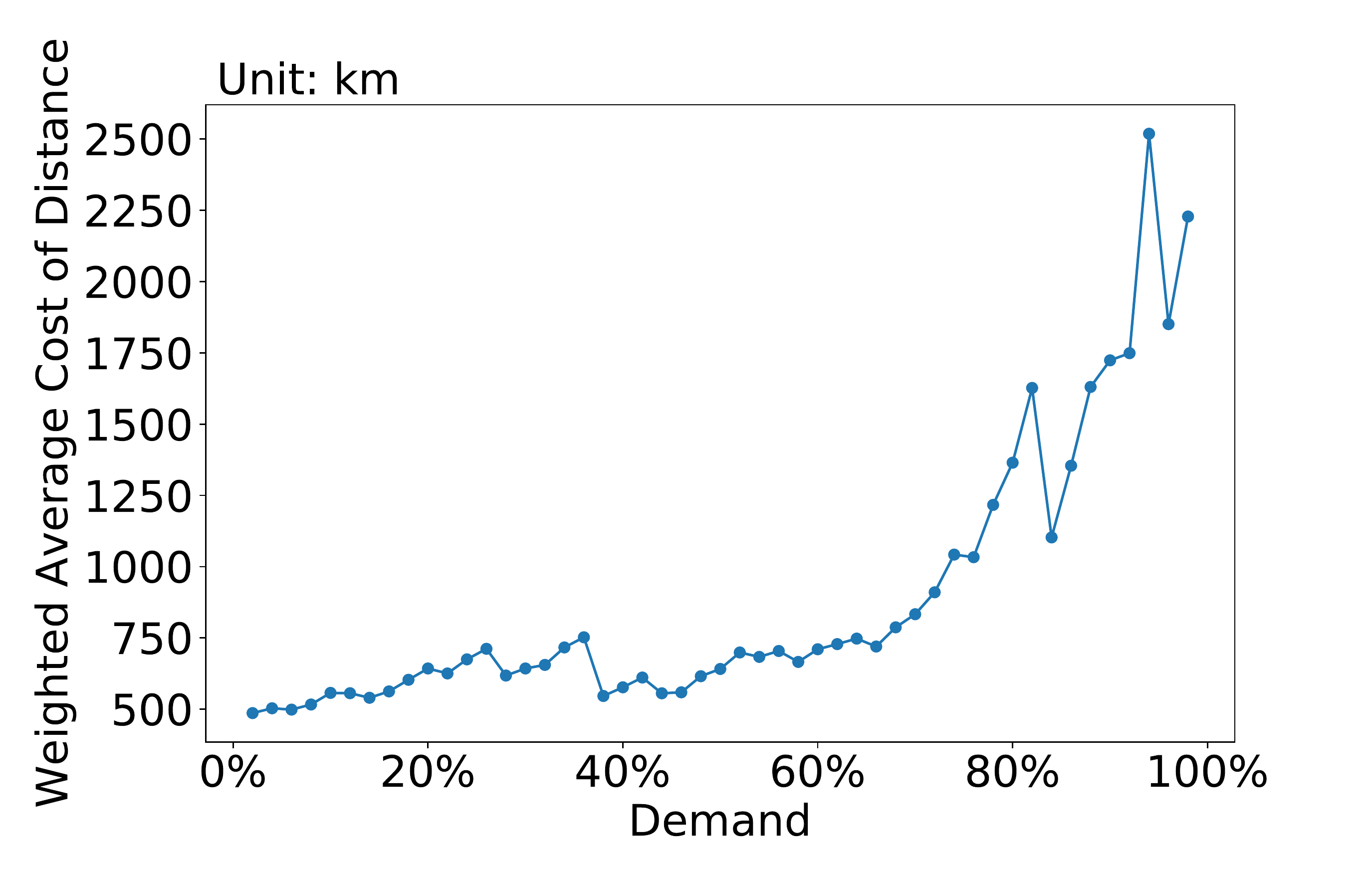}
  \includegraphics[width=0.45\textwidth]{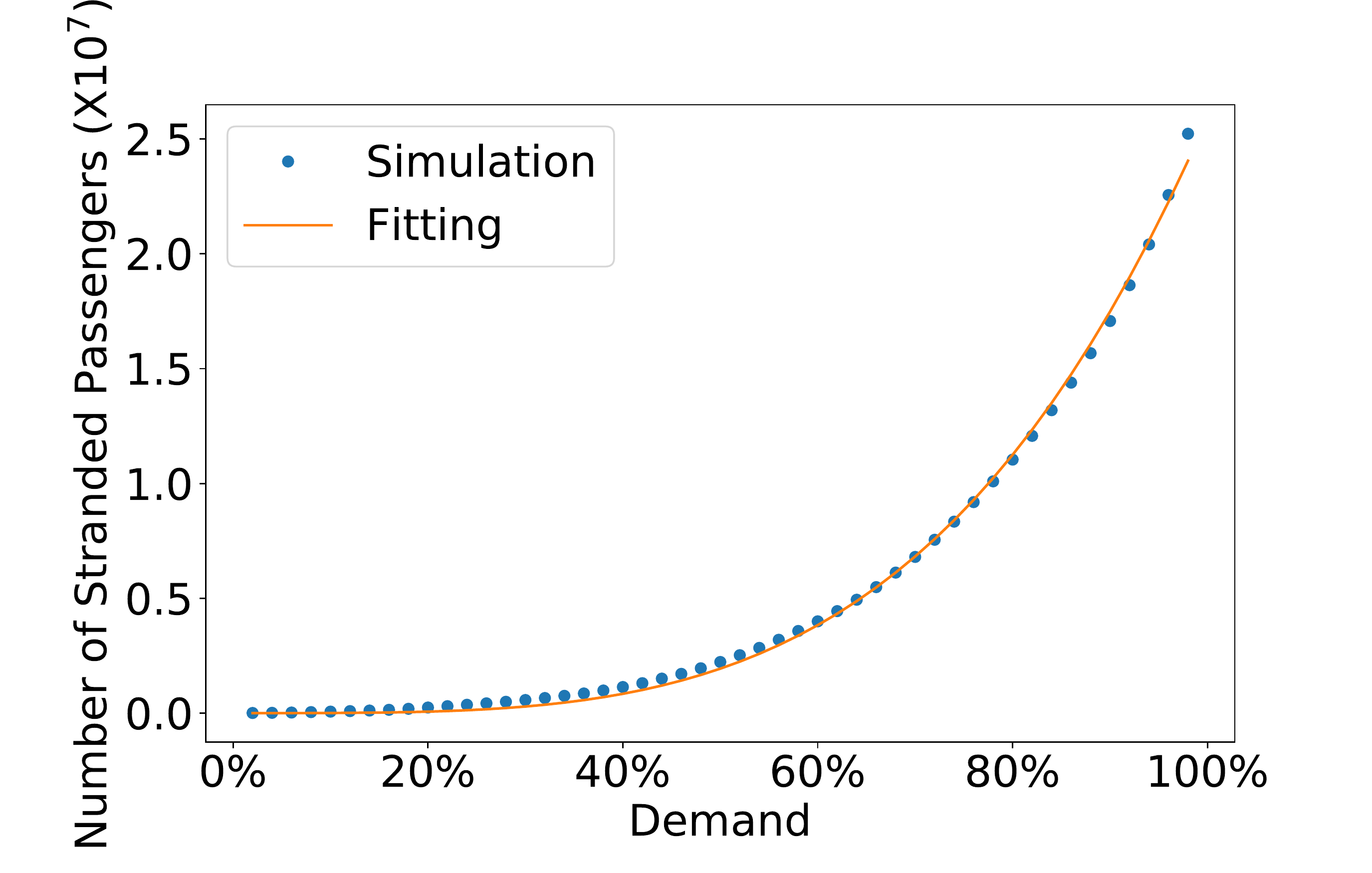}
  \includegraphics[width=0.45\textwidth]{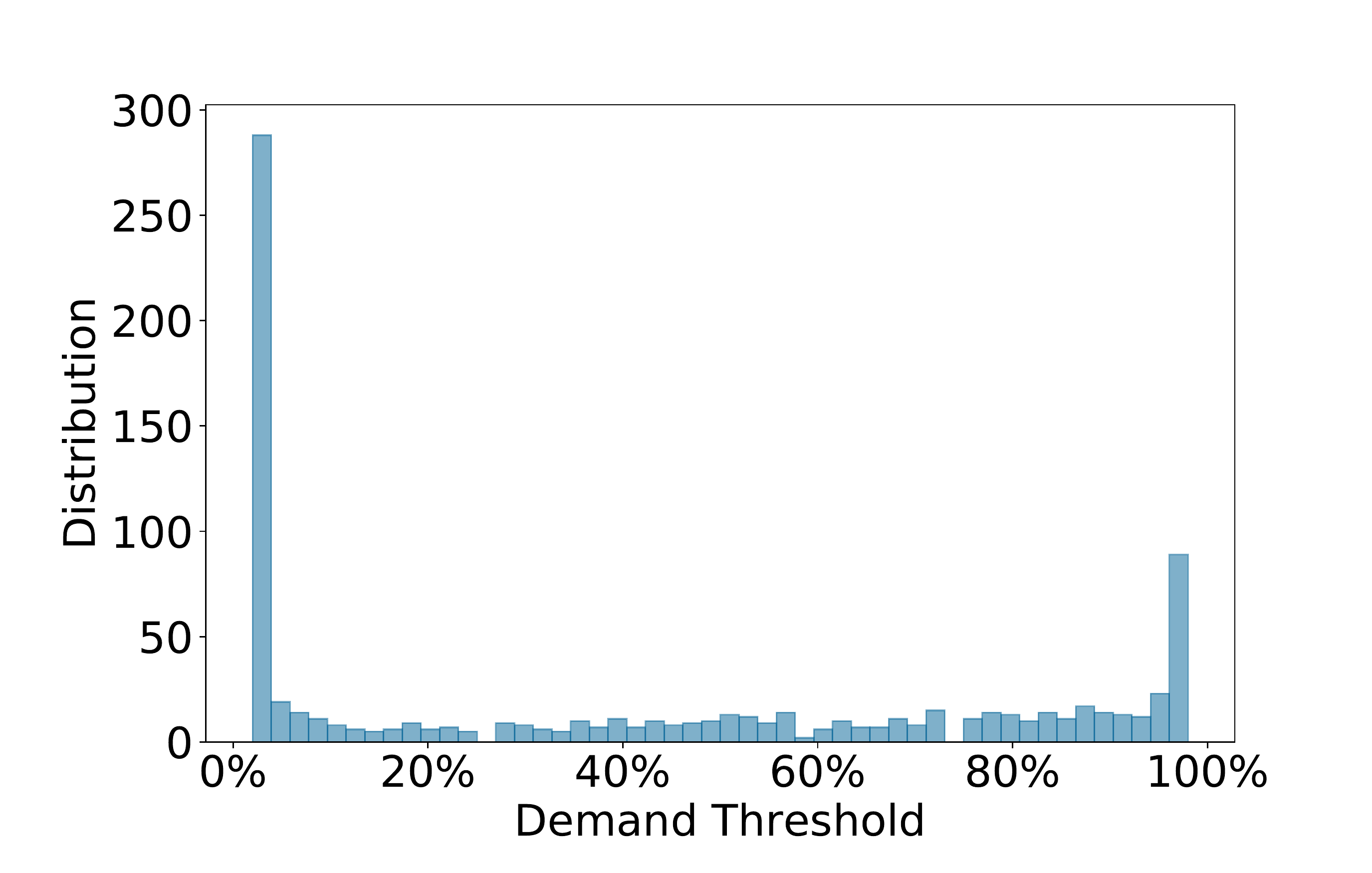}
  \caption{Observations on passenger rerouting after airline company bankruptcy. (a) Weighted average cost of additional distance with weight given by the capacity.  (b) Total number $N_s$ of stranded passengers as a function of $\alpha$ (the solid line indicates a power law fit of the form $N_s\propto\alpha^\tau$ where the exponent is here $\tau=3.75$). (c) Distribution of demand threshold.}
  \label{Rerouting}
\end{figure}

\section{Local Effective Demand}

The local effective demand \(E_i^s(\alpha)\) is defined as
\begin{equation}
E_i^s(\alpha) = \frac{N^s_i(\alpha)}{N_i},
\end{equation}
where $N^s_i(\alpha)$ represents the actual number of passengers for company \(i\) when passengers follow the airline choice strategy \(s\). Small values of this quantity allow us to identify companies at risk of bankruptcy.

\begin{figure}[h]
  \centering
  \includegraphics[width=0.5\textwidth]{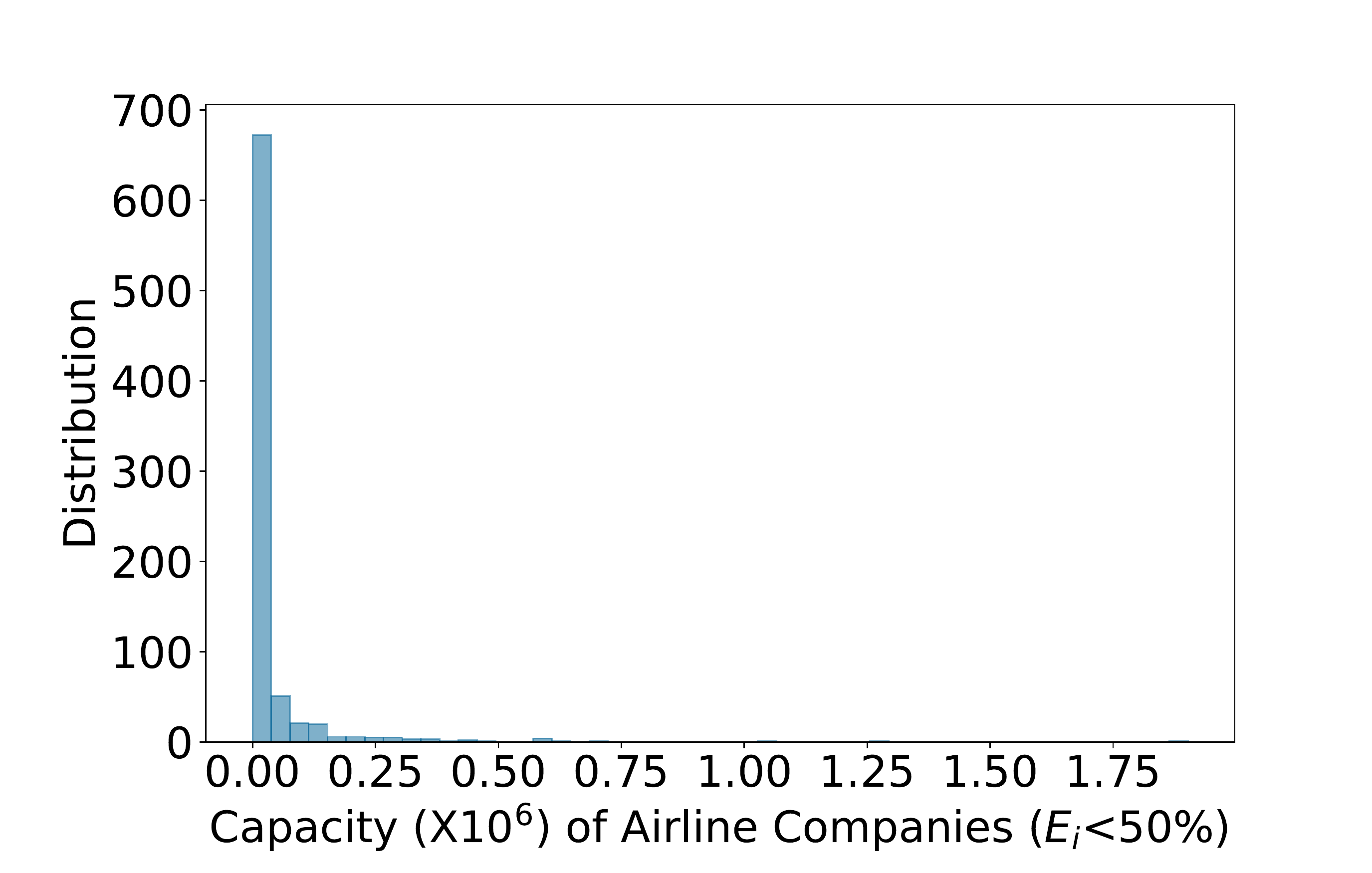}
  \caption{Distribution of capacity of companies with local effective demand less than $50\%$ when \(\alpha=36\%\), the estimated value of real demand in April 2020.}
  \label{SI_EffectiveDemand_Capacity}
\end{figure}

We first show in Fig.~\ref{SI_EffectiveDemand_Capacity}, for the strategy of `Biggest First', the distribution of the capacity of the airlines companies with local effective demand less than $50\%$ (computed for the global demand \(\alpha=36\%\) which is the estimated value of real demand in April 2020 by IATA). We see that even if most companies meeting this condition have a small capacity, there are however a few large ones.

\begin{figure}[h]
  \centering
  \includegraphics[width=0.45\textwidth]{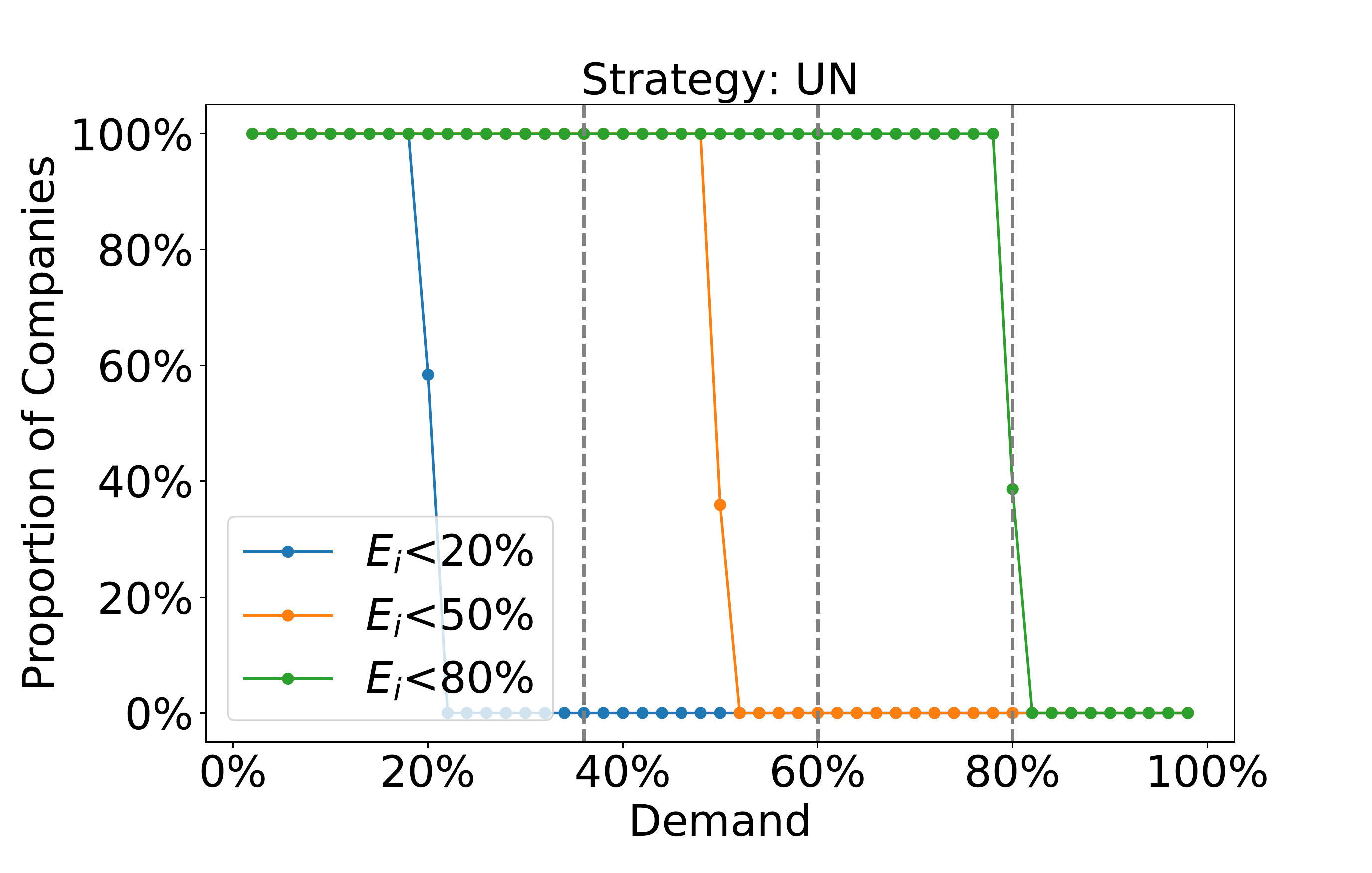}
  \includegraphics[width=0.45\textwidth]{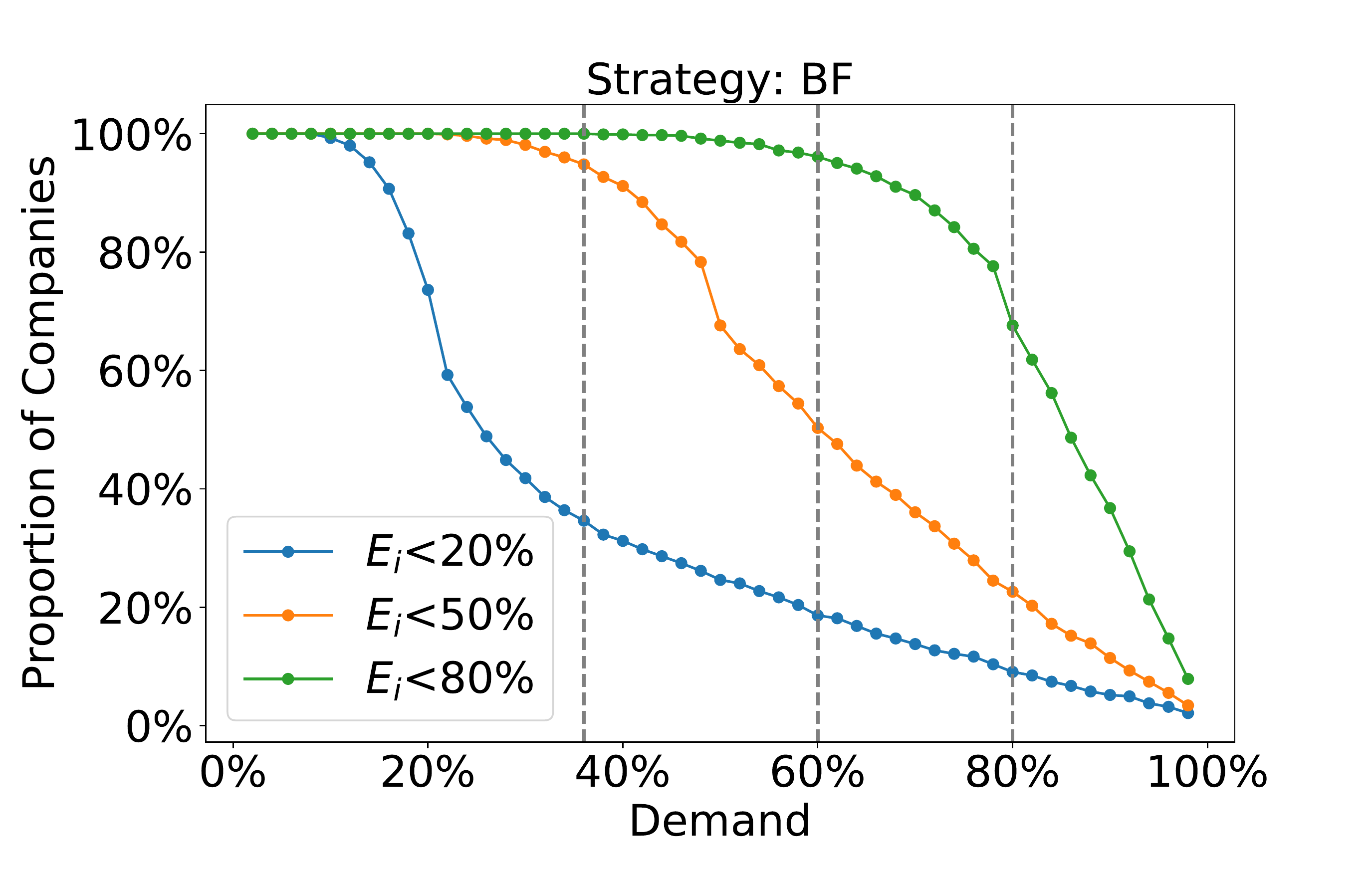}
  \includegraphics[width=0.45\textwidth]{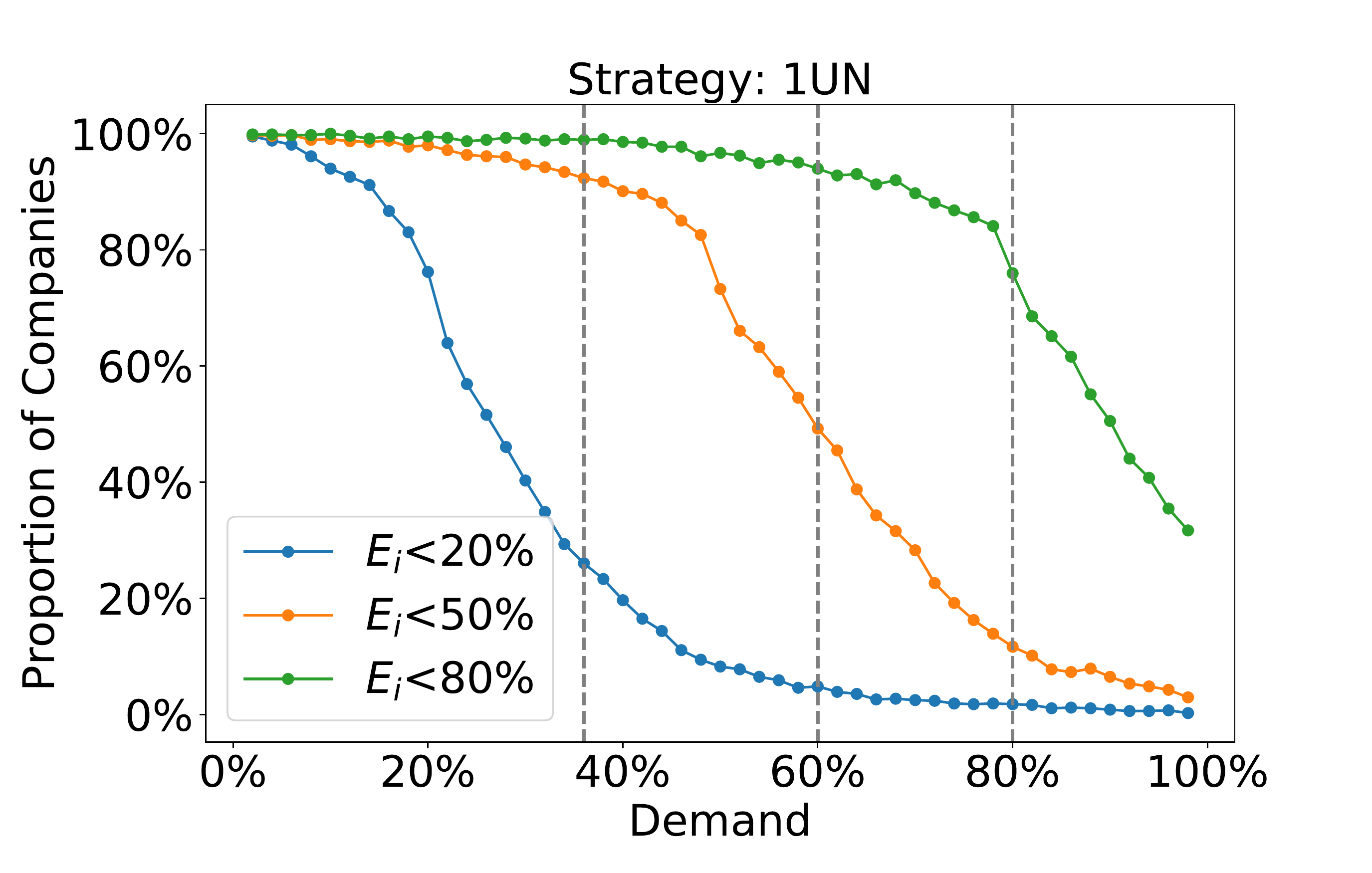}
  \includegraphics[width=0.45\textwidth]{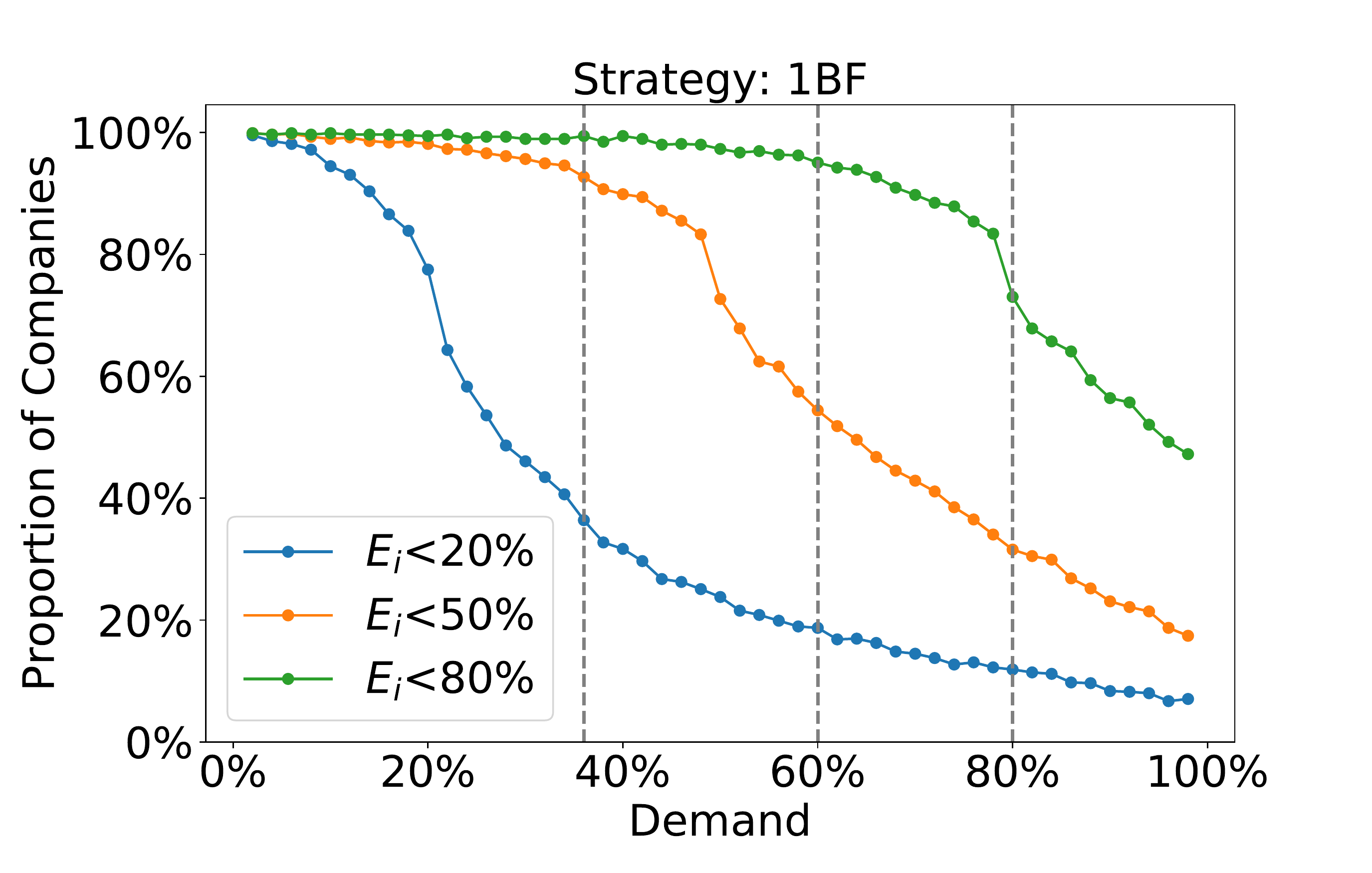}
  \includegraphics[width=0.45\textwidth]{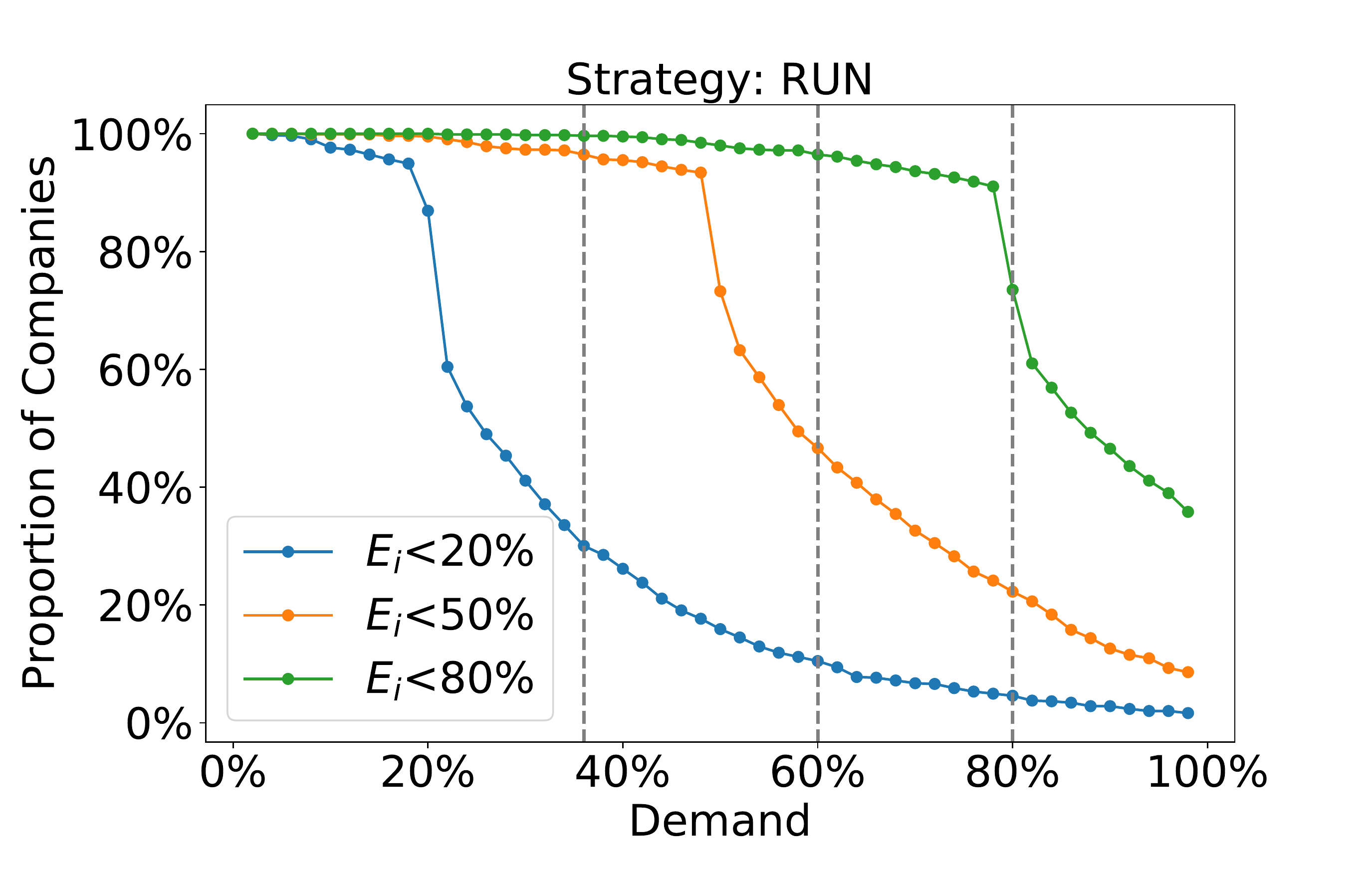}
  \includegraphics[width=0.45\textwidth]{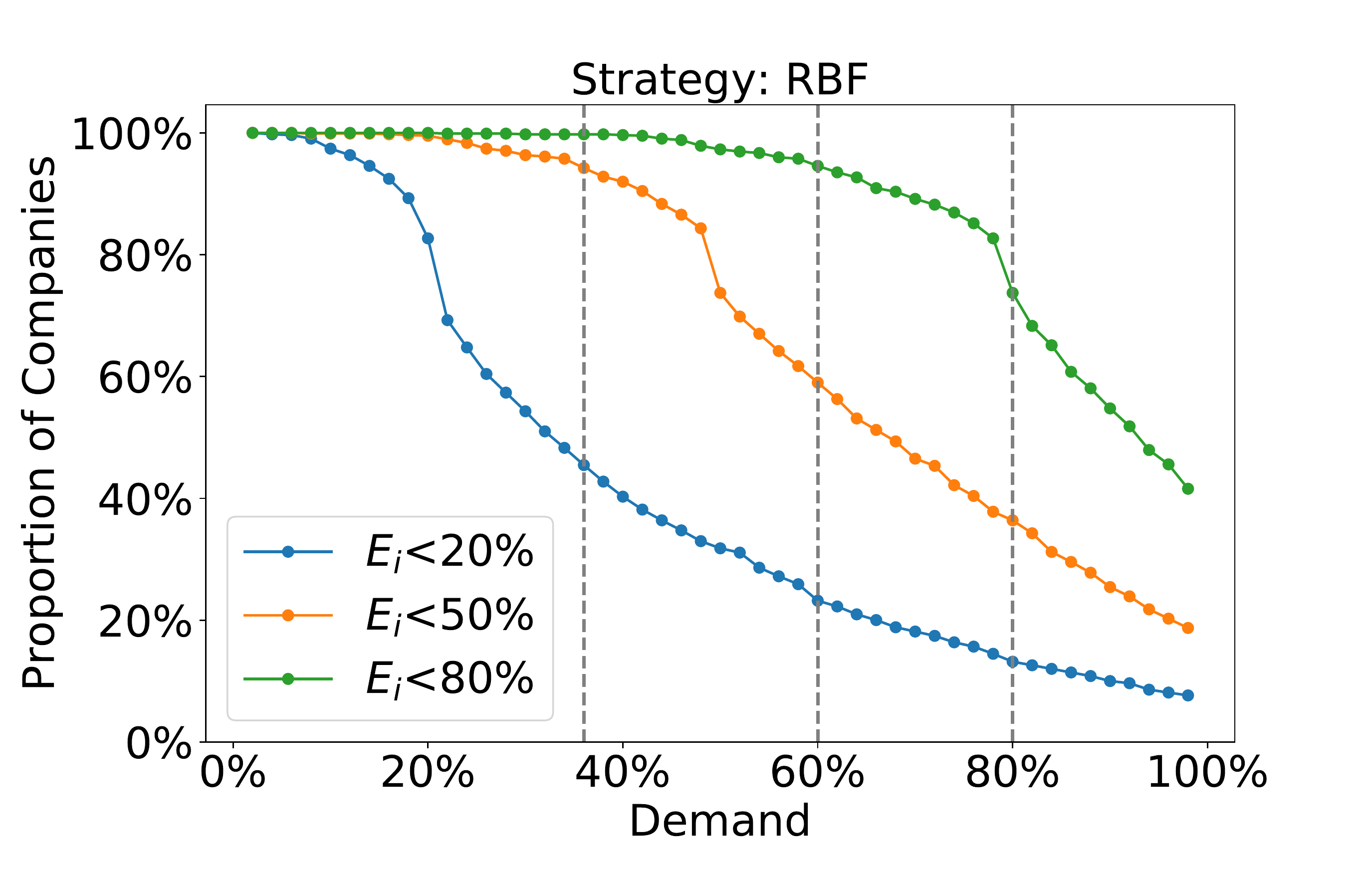}
  \caption{Proportion of companies with local effective demand less than different threshold. For each strategy, three thresholds are compared: $0.2$, $0.5$ and $0.8$. (a) `Uniform' strategy. (b) `Biggest First' strategy. (c) `One-Uniform' strategy. (d) `One-Biggest First' strategy. (e) `Rank-Uniform' strategy. (f) `Rank-Biggest First' strategy. The gray dotted lines highlight \(\alpha=36\%\), $60\%$ and $80\%$.}
  \label{SI_EffectiveDemand_Proportion}
\end{figure}

\begin{figure}[t]
  \centering
  \includegraphics[width=0.5\textwidth]{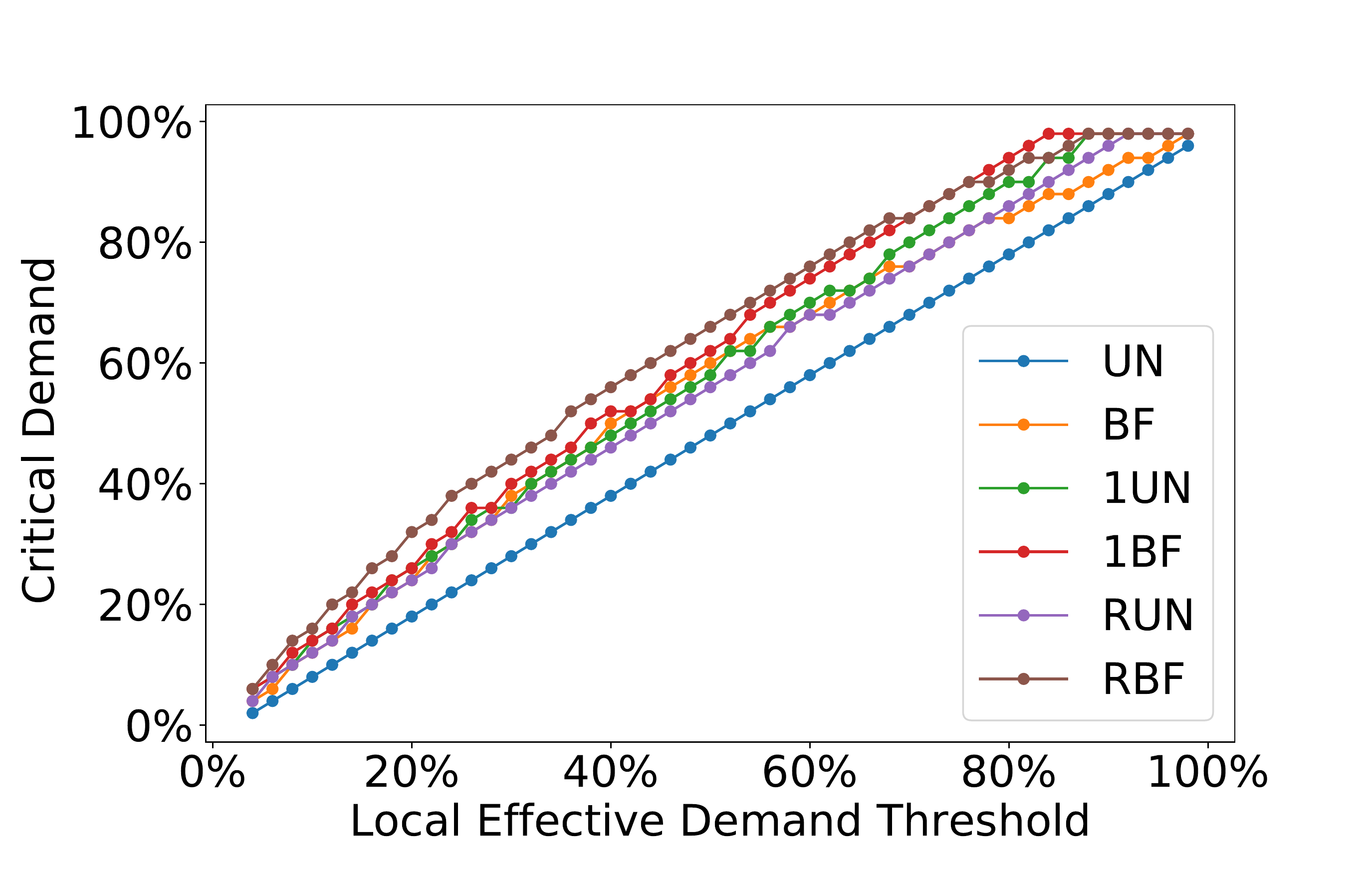}
  \caption{Comparison of critical demand as the function of local effective demand threshold between different strategies.}
  \label{SI_EffectiveDemand_Critical}
\end{figure}

For each strategy, we compare the proportion of companies with local effective demand less than a value equal to different thresholds $20\%$, $50\%$ or $80\%$ (Fig.~\ref{SI_EffectiveDemand_Proportion}). We observe a similar behavior for all these different thresholds. In addition, for a given threshold, there is a critical point for $\alpha$, below which the proportion of companies is above $50\%$. In other words, for \(\alpha\) less than this critical point, the proportion of companies with local effective demand less than the threshold is larger than $50\%$, while for \(\alpha\) larger than the critical point, the proportion of companies with local effective demand less than the threshold is less than $50\%$. We show the value of the critical point of demand as a function of the demand threshold and for different strategies in Fig.~\ref{SI_EffectiveDemand_Critical}. We observe that this critical point is always larger than the threshold, meaning that in general the local effective demand is always smaller than the global one.

\end{document}